\documentclass[aps,pra,twocolumn]{revtex4}
\usepackage{graphicx}

\usepackage{bm}

\newcommand{\br}[1]{\langle #1|}
\newcommand{\ke}[1]{|#1\rangle}

\newcommand{\kb}[2]{\ke{#1}\br{#2}}
\newcommand{\da}{^\dagger}
\newcommand{\pt}[1]{\left( #1 \right)}
\newcommand{\pq}[1]{\left[ #1 \right]}
\newcommand{\pg}[1]{\left\{ #1 \right\}}

\newcommand{\av}[1]{\left\langle #1 \right\rangle}

\begin{document}

\title{Mechanical effects of optical resonators on driven trapped atoms: \\
Ground state cooling in a high finesse cavity}
\author{Stefano Zippilli$^1$}
\author{Giovanna Morigi$^2$}
\affiliation{$^1$ Abteilung f\"ur Quantenphysik,
Universit\"at Ulm, D-89069 Ulm, Germany\\
$^2$ Grup d'Optica, Departament de Fisica, Universitat Aut\`onoma de Barcelona, 08193 Bellaterra, Spain}

\date{\today}
\begin{abstract}
We investigate theoretically the mechanical effects of light on
atoms trapped by an external potential, whose dipole transition
couples to the mode of an optical resonator and is driven by a
laser. We derive an analytical expression for the quantum
center-of-mass dynamics, which is valid in presence of a tight
external potential. This equation has broad validity and allows for a transparent
interpretation of the individual scattering processes leading to cooling. We
show that the dynamics are a competition of the mechanical effects
of the cavity and of the laser photons, which may mutually
interfere. We focus onto the good-cavity limit and identify novel cooling schemes, which are based on quantum interference effects and lead to efficient
ground state cooling in experimentally accessible parameter
regimes.
\end{abstract}
\maketitle

\section{Introduction}

Atom cooling by photon scattering is achieved by enhancing the
rate of scattering processes that dissipate motional energy,
thereby exploiting the conservation of internal and mechanical
energy in the interaction between atoms and electromagnetic
field~\cite{ReviewCooling}. The atomic scattering cross section
can be significantly modified by the coupling to an optical
resonator, which acts both on the internal as well as on the
external degrees of freedom. Hence, the scattering properties can
be tailored, allowing to achieve efficient cooling also for atoms
and molecules which may not offer a convenient configuration in
free space~\cite{Vuletic00,Domokos03}. This principle is at the
basis of cooling by means of an optical resonator. Indeed, the
mechanical effects on atoms coupled to an optical resonator are
object of several
experimental~\cite{Pinkse00,Hood00,Rempe04,KimbleFORT03,Vuletic03,Zimmermann03,Hemmerich03,Buschev04,Kuhn05}
and
theoretical~\cite{Domokos03,Cirac95,Horak,Vuletic01,vanEnk,Domokos02,Domokos02b,Domokos04,Beige04,Helmut04,Karim}
investigations, which aim at developing a systematic understanding
of these complex dynamics both for its fundamental aspects, as
well as for the perspective of a high degree of control of complex
systems with scalable number of degrees of freedom.

\begin{figure}[h]
\includegraphics[width=6cm]{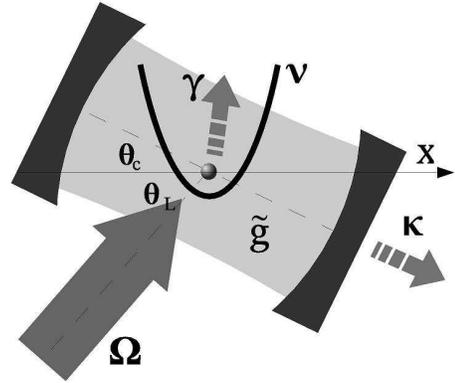}
\caption{An atom is confined by an harmonic potential of frequency $\nu$ inside an
optical resonator. A mode of the resonator couples with strength $\tilde g$  to the dipole,
which is driven transversally by a laser at Rabi frequency $\Omega$. The system dissipates by
spontaneous emission of the atomic excited state at rate $\gamma$
and by cavity decay at rate $\kappa$. The other parameters are discussed in Sec.~\ref{Sec:Model}.} \label{Fig:1}
\end{figure}

In this work we investigate the cooling dynamics of atoms inside
optical resonators, when their center-of-mass motion is tightly
confined by an external potential, like for instance a
dipole~\cite{KimbleFORT03,Sauer03} or an ion
trap~\cite{Guthorlein01,Keller04,Mundt02,Buschev04}. We consider
the situation where an atomic optical dipole transition is driven
by a laser and by a cavity resonator, as sketched in
Fig.~\ref{Fig:1}, and discuss in detail the results presented
in~\cite{Zippilli05}. In particular, we show the detailled
derivation of the rate equation discussed in~\cite{Zippilli05}.
This equation has broad validity, which is supported by numerical
checks, and allows for a transparent interpretation of the
individual scattering processes leading to cooling. Moreover, in
the corresponding parameter regimes it reproduces the results
reported in~\cite{Cirac95} and~\cite{Vuletic01}.

In this manuscript we mostly focus onto the good-cavity limit. In
this regime we discuss when efficient cooling into the potential
ground state can be achieved. In particular, we show that in
experimentally accessible parameter regimes one may obtain almost
unit ground state occupation, even when the natural linewidth of
the dipole transition would not allow for ground state cooling in
free space. Efficient ground-state cooling is often found by
exploiting interference effects, arising from phase correlation
between the laser and the field scattered by the atom into the
resonator. Most of these interference effects are due to the
discrete nature of the spectrum of the center-of-mass motion,
which is trapped by a harmonic potential. Hence, the dynamics here
studied differ substantially from the ones of cooling of free
atoms inside
cavities~\cite{Domokos02,Horak,Domokos02b,Domokos03,Domokos04}.
Such interference effects are at the basis of novel cooling
schemes, some of which have been identified in~\cite{Zippilli05},
and which are discussed in detail in the present work.

This article is organized as follows. In Sec.~\ref{Sec:II} some
preliminary considerations are made. In Sec.~\ref{Sec:Model} the
model is introduced, and the basic equations for the motion are
obtained. In Sec.~\ref{Sec:III} we discuss the dynamics of cooling
from the rate equation we obtain and review previous results
presented in the literature. In Sec.~\ref{Sec:4:Interfere} novel
cooling schemes are presented, whose dynamics are due to quantum
correlations which are established in the good cavity limit.  In
Sec.~\ref{Sec:Results} the results are reported: The cooling
efficiencies in the various parameter regimes are discussed and
compared. In Sec.~\ref{Sec:Conclusions} the conclusions are drawn.
The appendices report detailed calculations at the basis of the
equations derived in Secs.~\ref{Sec:Model},~\ref{Sec:III}
and~\ref{Sec:4:Interfere}.

\section{Mechanical effects of cavity and laser on the atomic motion}
\label{Sec:II}

\begin{figure}[h]
\includegraphics[height=5cm]{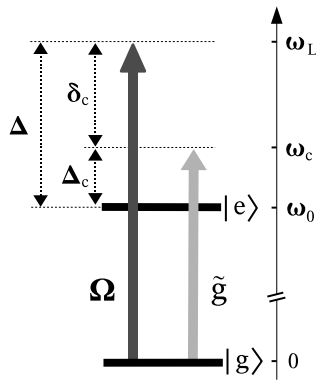}
\caption{ Sketch of the
internal levels $\ke{g}$ and $\ke{e}$ of the atomic dipole
transition, driven by a laser and a cavity mode with coupling strengths $\tilde{g}$ and $\Omega$, respectively. The arrows show the cavity and laser
frequency with respect to the dipole frequency. Here, $\Delta$ and
$\delta_c$ are the detunings of atom and cavity, respectively,
from the laser frequency, $\Delta_c$ is the detuning of the atomic transition  from the cavity frequency. The frequency of the atomic transition ($\omega_0$), of the cavity mode ($\omega_c$) and of the laser field ($\omega_L$) are indicated in the vertical scale.} \label{Fig:1b}
\end{figure}

In this section we make some physical considerations, in order to provide insight into the results presented in the rest of the manuscript. The scattering cross section of the bare atom is usually very informative about the cooling process~\cite{Eschner03}. When the atomic transition is driven in saturation, this analysis is more conveniently done in the dressed state picture~\cite{Dalibard85,Morigi00,Domokos04}. At this purpose, we first
assume that the atom is fixed at the position $x$, such that the coupling constant to the cavity mode is $\tilde g=g(x)$. The dipole level scheme and the relevant parameters are shown in Fig.~\ref{Fig:1b}. We denote by $|g,n_c\rangle$ and $|e,n_c\rangle$ the states of the system,
where $\ke{g}$, $\ke{e}$ are the ground and excited state of the
atomic dipole and $|n_c\rangle$ the number of photons of the
cavity mode. In the situation in which the atom is strongly coupled to the cavity mode and weakly pumped by the laser, the states which are relevantly involved into the dynamics are $|g,0_c\rangle$ and the dressed states
\begin{eqnarray}
&&|+\rangle=\sin\vartheta |g,1_c\rangle + \cos\vartheta |e,0_c\rangle
\\
&&|-\rangle=\cos\vartheta |g,1_c\rangle -\sin\vartheta |e,0_c\rangle
\end{eqnarray}
with $$\tan\vartheta=\tilde g/(-\Delta_c/2+\sqrt{\tilde g^2+\Delta_c^2/4})$$ and
$\Delta_c$ the detuning between cavity mode and atom. Setting
the energy of $|g,0_c\rangle$ at zero, the frequencies of the
states $|\pm\rangle$ are
\begin{eqnarray}
\lambda_{\pm}=-\Delta_c/2\pm \sqrt{\tilde g^2+\Delta_c^2/4}
\end{eqnarray}
and the respective linewidths are $\gamma_+\sim\kappa\sin^2\vartheta+\gamma\cos^2\vartheta$,
$\gamma_-\sim\kappa\cos^2\vartheta+\gamma\sin^2\vartheta$,
where $\gamma$ is the linewidth of the dipole transition and
$\kappa$ the cavity decay rate. The weak laser probe couples the dressed states $|g,0_c\rangle$ and $|e,0_c\rangle$.

\begin{figure}[h]
\includegraphics[width=8cm]{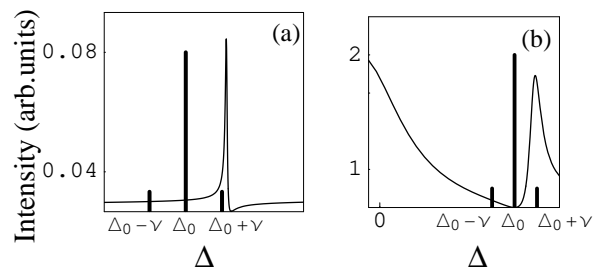}
\caption{Excitation spectrum as a function of the laser detuning
$\Delta$ in the reference frame of the atom. Here, $\tilde
g=0.5\gamma$, $\kappa=0.01\gamma$. In (a) $\Delta_c=-10\gamma$ and
(b) $\Delta_c=1.2\gamma$. The vertical bars indicate the frequency
$\Delta_0$ of the carrier (central line) and of the red and blue
sideband transitions, $\Delta_0+\nu$ and $\Delta_0-\nu$,
respectively, when the laser is set at $\Delta=\Delta_0$ and the
trap frequency $\nu=0.2\gamma$. In (a) $\Delta_0\sim\Delta_c-\nu$;
In (b) $\Delta_0=\Delta_c$, namely cavity mode and laser are
resonant. See text.} \label{Fig:2}
\end{figure}

Signatures of the dressed states are for instance the resonances
in the rate of photon scattering as obtained by scanning the probe
laser through atomic resonance. This situation is depicted in
Fig.~\ref{Fig:2}. Here, the curve has been evaluated for a good
resonator, namely $\kappa\ll\gamma,\tilde g$ and $\Delta_c\neq0$.
For these parameters the linewidth of one of the two resonances is
narrower than the natural linewidth of the dipole. Moreover, when
the probe laser is resonant with the cavity mode, the spectrum
exhibits a minimum, which reaches zero for $\kappa=0$, namely no
photons are scattered. This behaviour is due to an interference
effect between laser and cavity resonator, such that there is no
radiation scattered by the atom, as it is at a point where the two
fields, laser and cavity, mutually
cancel~\cite{Alsing92,Zippilli04a,Zippilli04b}.

We now consider the center-of-mass motion of an atom in a harmonic
oscillator, and first assume that the mechanical effects are only
due to the laser, while the cavity wave vector is orthogonal to
the motional axis. In this regime, the motion gives rise to a
modulation of the laser frequency at the trap frequency $\nu$. In
the regime of strong confinement (Lamb-Dicke regime) this gives
rise to two sidebands of the carrier, i.e.\ the laser frequency.
The carrier and sideband positions are indicated by the vertical
bars in Fig.~\ref{Fig:2} in the reference frame of the atom. The
central bar is the carrier. The bar at the right (left) of the
carrier corresponds to a transition which lowers (rises) the
atomic vibrational excitation by one phonon, namely the so-called
red (blue) sideband transition. These two components are out of
phase with respect to the carrier. In the limit in which the
atomic motion weakly perturbs the internal and cavity dynamics,
the scattering along the sidebands is proportional to the
corresponding value of the excitation spectrum. Cooling is thus
obtained by realizing a large gradient between scattering rates
along the sidebands. Figures~\ref{Fig:2}(a) and (b) show two
possible scenarios, which are discussed in this paper. Case (a)
corresponds to use the narrow resonance for implementing sideband
cooling with the dressed states~\cite{Vuletic01}. This scenario is
obtained by choosing a large value of $|\Delta_c|$ and setting the
detuning between the cavity and the laser equal to the trap
frequency, such that the red sideband absorption falls at the
center of the narrow resonance. This case has been studied
in~\cite{Vuletic01}. In case (b) a large gradient is achieved by
exploiting the interference profile arising when laser and cavity
are resonant.

\begin{figure*}[t]
\includegraphics[width=18cm]{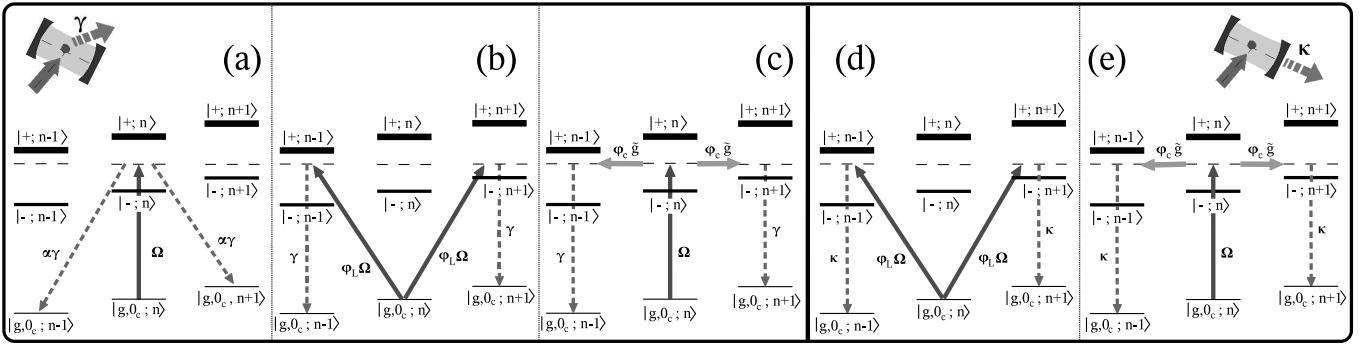}
\caption{Scattering processes leading to a change of the
vibrational number by one phonon. The states $|g,0_c;n\rangle$,
$\ke{\pm;n}$ are the cavity-atom dressed states at phonon number
$n$. Processes (a),(b),(c) describe scattering of a laser photon
by spontaneous emission. They prevail in good resonators, for
$\kappa\ll\tilde g,\gamma$. Processes (d)
and~(e) describe scattering of a laser photon by cavity decay.
They prevail in bad resonators, for $\gamma\ll\tilde g,\kappa$. The parameters $\alpha,\varphi_c,\varphi_L$
emerge from the mechanical effects of light and are defined in Sec.~\ref{rateeq}.}
\label{Fig:3}
\end{figure*}

The dressed state picture, as obtained by neglecting the motion of
the atom, can be also applied to get some insight into the cooling
dynamics when cooling is due only to the resonator forces or to
both laser and resonator. Nevertheless, it does not explain other
cooling dynamics, which we discuss in this article, and which are
due to correlations in the gradients of the fields over the atomic
wave packet. At this purpose one has to consider also the quantum
motion.

Figure~\ref{Fig:3} summarizes the basic scattering processes
determining the cooling dynamics in the basis $|g,0_c;n\rangle$,
$\ke{\pm;n}$, where $n$ is the number of excitations of the
center-of-mass harmonic oscillator. The process shown in
Fig.~\ref{Fig:3}(a) describes absorption of a laser photon and
spontaneous emission, whereby the change in the center-of-mass
state is due to the recoil induced by the spontaneously emitted
photon. The scattering rate is scaled by the geometric factor
$\alpha$ and is found after averaging over the solid angle of
photon emission into free space. This contribution is diffusive,
as the motion can be scattered into a higher or lower vibrational
state with probabilities depending on the overlap integrals
between the initial and the final motional states after a photon
recoil. In section~\ref{Sec:4:Interfere} we discuss the parameter
regime in which this contribution can be suppressed by an
interference effect in the dressed states absorption.

The processes depicted in Figs.~\ref{Fig:3}(b) and~(c) describe scattering of a laser photon by
spontaneous emission where the motion is changed by mechanical
coupling to the laser,(b), and to the cavity,(c), field. Since the final
state of the two scattering processes is the same, they
interfere. In addition, each term is composed by
multiple excitation paths, and can vanish in some parameter
regimes. In section~\ref{Sec:4:Interfere} we discuss interference effects in these
two terms.

The processes depicted in Figs.~\ref{Fig:3}(d) and~(e) describe scattering of a laser photon by
cavity decay, where the motion is changed by mechanical coupling
to the laser,(d), and to the cavity,(e), field. The scattered photon is transmitted
through the cavity mirrors into the external modes, and therefore
these two processes do not interfere with the ones above
discussed, but add up coherently with one another. In section~\ref{Sec:4:Bad} we discuss
parameter regimes where interference effects in these two terms relevantly affect the dynamics.
The general dynamics are a competition of all these processes, and will be
discussed in detail in the following sections.

\section{The Model}
\label{Sec:Model}

\subsection{Basic Equations}

We consider an atom of mass $M$, which is confined by a harmonic
potential of frequency $\nu$ inside an optical resonator. The relevant center-of
-mass
dynamics are along the $x$-axis, while the degrees of freedom of
the transverse motion have been traced out, assuming that the
transverse confinement is much steeper. Later on we discuss how the treatment can be
generalized to three dimensional motion. The atom internal degrees
of freedom, which are relevant to the dynamics, are the ground
state $\ke{g}$ and the excited state $\ke{e}$, constituting a
dipole transition at frequency $\omega_0$ and linewidth $\gamma$.
The dipole couples with a cavity mode at frequency $\omega_c$ and
with a laser at frequency $\omega_L$, whose wave vectors ${\bf k_c}$ and
${\bf k_L}$ form the angle $\theta_c$ and $\theta_L$, respectively, with the $x$-axis.
The system is sketched in Fig.~\ref{Fig:1} and~\ref{Fig:1b}.
We denote by $\rho$ the density matrix for the atom and the
resonator degrees of freedom in the reference frame rotating at
the laser frequency. The density matrix $\rho$ obeys the master equation
\begin{eqnarray}\label{meq}
\frac{\partial}{\partial t}\rho=
\frac{1}{{\rm i}\hbar}[H,\rho]+{\cal L}_s\rho+{\cal K}\rho\equiv{\cal L}\rho
\end{eqnarray}
where ${\cal L}$ is the Liouvillian describing the total dynamics.
Here, the Hamiltonian $H$ is
\begin{eqnarray}
\label{H:tot}
H=H_{\rm mec}+H_{\rm at}+H_{\rm cav}+H_{\rm at-cav}+H_L
\end{eqnarray}
where the terms describing the coherent dynamics in absence of
coupling with the e.m.-field are
\begin{eqnarray}
\label{Hmec}
&&H_{\rm mec}=\frac{p^2}{2 M}+\frac{1}{2}M\nu^2 x^2\\
&&H_{\rm at}=-\hbar\Delta\sigma\da\sigma\\
&&H_{\rm cav}=-\hbar\delta_ca\da a
\end{eqnarray}
Here, $x$, $p$ are position and momentum
of the center of mass; $\sigma=\ke{g}\br{e}$, and $\sigma\da$ its
adjoint; $a$, $a\da$ are the annihilation and creation operators
of a cavity photon; $\Delta=\omega_L-\omega_0$ and
$\delta_c=\omega_L-\omega_c$ are the detunings of the laser from
the dipole and from the cavity frequency, respectively, such that
$$\Delta_c=\Delta-\delta_c.$$ The terms
\begin{eqnarray}
\label{Hat-cav}
&&H_{\rm at-cav}=\hbar g \cos(k x\cos\theta_c+\phi)(a\da\sigma+a\sigma\da)\\
&&H_L=\hbar\Omega(e^{{\rm i}k x\cos\theta_L}\sigma\da+{\rm H.c.})
\label{HL}
\end{eqnarray}
describe the radiative couplings of the dipole with the cavity
mode and the laser, respectively, where $g$ is the cavity-mode
vacuum Rabi frequency and $\Omega$ the Rabi frequency for the
coupling with the laser, $\phi$ is a phase, and $k$ is the modulus
of the wave vector ($|{\bf k_L}|\approx|{\bf k_c}|\approx \omega_0/c=k$).

The superoperators ${\cal K}$ and ${\cal L}_s$ in Eq.~(\ref{meq}) describe the
cavity decay and  dipole spontaneous emission into the modes
external to the resonator, respectively, and are
\begin{eqnarray}\label{atomdecay}
&&{\cal K}\rho=\frac{\kappa}{2}(2a\rho a\da-a\da a\rho-\rho
a\da a)\\
&&{\cal L}_s\rho=\frac{\gamma}{2}\left(
2\sigma\tilde{\rho}\sigma\da
-\sigma\da\sigma\rho-\rho\sigma\da\sigma\right)
\end{eqnarray}
where $\kappa$ is the cavity decay rate due to the finite
transmission at the mirrors and
\begin{eqnarray}
\tilde{\rho}=\int_{-1}^1 {\rm d}\cos\theta_0 {\cal N}(\theta_0)
e^{-{\rm i} k\cos\theta_0 x}\rho e^{{\rm i} k\cos\theta_0 x}
\end{eqnarray}
describes the events in which the atomic motion recoils
by emission of a photon at the angle $\theta_0$ with the trap
axis with probability ${\cal N}(\theta_0){\rm d}\cos\theta_0$.
Note that ${\cal N}(\theta_0)$ must be evaluated taking into
account the geometry of the setup.

For later convenience we introduce the annihilation and creation
operators $b$ and $b\da$ of a quantum of vibrational energy, such
that
\begin{eqnarray}
&&x=\sqrt{\hbar/2M\nu}(b\da+b)\label{x}\\
&&p={\rm i}\sqrt{\hbar M\nu/2}(b\da-b),
\end{eqnarray}
and the Hamiltonian term~(\ref{Hmec}) can be rewritten as
\begin{equation}
H_{\rm mec}=\hbar\nu\left(b\da b+\frac{1}{2}\right)
\end{equation}
We denote by $|n\rangle$ the eigenstates of $H_{\rm mec}$ at the
eigenvalue $(n+1/2)\hbar\nu$ and introduce the Lamb-Dicke
parameter
\begin{equation}
\eta=k\sqrt{\frac{\hbar}{2M\nu}}
\end{equation}
which scales the mechanical coupling of radiation with the atomic motion.

\subsection{Reduced equation for the center-of-mass dynamics in the Lamb-Dicke limit}\label{rateeq}

We assume the Lamb-Dicke regime, namely the atom is localized on a
length scale which is much smaller than the light wave length, and
identify in the Lamb-Dicke parameter $\eta$ the perturbative
parameter, that allows us to treat the coupling of the external
degrees of freedom with the cavity and the atom internal degrees
of freedom in perturbation theory~\cite{Stenholm86,Javanainen84}.
We apply the formalism first applied in~\cite{Javanainen84} and
then further developed in~\cite{Cirac92,Morigi03,Bienert04}. Below
we summarize some steps.

At zero order in the Lamb-Dicke parameter the center-of-mass is
decoupled from the internal and cavity degrees of freedom. In
fact, denoting by ${\cal L}_0={\cal L}|_{\eta=0}$ the Liouvillian
at zero order in the expansion, this can be decomposed into the sum of a term acting over the external and over the cavity and dipole degrees of freedom, namely $${\cal L}_0={\cal
L}_{0E}+{\cal L}_{0I},$$ where
\begin{eqnarray}
{\cal L}_{0E}\rho&=&\frac{1}{{\rm i}\hbar}[H_{\rm mec},\rho]\\
{\cal L}_{0I}\rho&=&\frac{1}{{\rm i}\hbar}[H_{\rm at}+H_{\rm cav}
+H_{0{\rm at-cav}}+H_{0L},\rho]\nonumber\\
&&+{\cal K}\rho+{\cal L}_{0s}\rho\label{Liouv0I}
\end{eqnarray}
and where the Hamiltonian interaction
\begin{eqnarray}\label{H0at-cav}
&&H_{0{\rm at-cav}}=\hbar \tilde{g}(a\da\sigma+a\sigma\da)\\
&&H_{0L}=\hbar\Omega(\sigma\da+\sigma)\label{H0L}
\end{eqnarray}
and the Liouvillian for the atomic spontaneous emission
\begin{eqnarray}
{\cal L}_{0s}\rho=\frac{\gamma}{2}\left(
2\sigma\rho\sigma\da
-\sigma\da\sigma\rho-\rho\sigma\da\sigma\right)
\end{eqnarray}
appear at zero order in the expansion in $\eta$. The term
\begin{equation}\label{tilde:g}\tilde g=g
\cos\phi\end{equation}
is the zero order atom--cavity coupling strength.

The spectrum of ${\cal L}_0$ is $\lambda=\lambda_I+\lambda_E$,
where $\lambda_I$ are the eigenvalues of ${\cal L}_{0I}$ and
$\lambda_E$ are the eigenvalues of ${\cal L}_{0E}$. The stationary
state is a right eigenstate at eigenvalue zero, as  it fulfills
the secular equation ${\cal L}_0\rho=0$~\cite{Englert}. The
corresponding eigenspace is spanned, for instance, by the
eigenvectors $\rho_n=\rho_{\rm St}\otimes \ke{n}\br{n}$, where
$\rho_{\rm St}$ fullfills the equation
$${\cal L}_{0I}\rho_{\rm St}=0$$
while the operator $\ke{n}\br{n}$ is eigenvector of the
superoperator ${\cal L}_{0E}$ at eigenvalue $\lambda_E=0$. The
corresponding eigenspace is infinitely degenerate. We denote by
$P$ the projection operator over the $\lambda=0$ eigenspace,
defined as
\begin{eqnarray}
P\rho=\rho_{\rm St}\otimes  \sum_{n=0}^\infty
\kb{n}{n}\mbox{Tr}_I\pg{\br{n}\rho\ke{ n}}
\end{eqnarray}
where $\mbox{Tr}_I$ is the trace over the dipole and cavity
degrees of freedom. At second order in $\eta$ one gets a closed
equation for the center of mass dynamics of the form
\begin{eqnarray}\label{externaleq}
\frac{d}{dt}\mu&=&\eta^2[ (S(\nu)+D)(b\mu b\da-b\da b\mu)\nonumber\\
&&+(S(-\nu)+D)(b\da\mu b-bb\da\mu)+{\rm H.c.}]
\end{eqnarray}
where $\mu={\rm Tr}_I\pg{P\rho}$ is the density matrix for the
center-of-mass variables, obtained by tracing over the dipole and
cavity degrees of freedom, and where the coefficients are given by
\begin{eqnarray}
\label{Diff}
D&=&\alpha\frac{\gamma}{2}\mbox{Tr}_I\{\sigma\da
\sigma\rho_{\rm St}\}\\
S(\nu)&=&\frac{1}{\hbar^2}\int_0^\infty d\tau e^{{\rm i}\nu\tau} \mbox{Tr}_I\pg{
V_1 e^{\mathcal{L}_{0I}\tau}V_1\rho_{ss}}\nonumber\\
&=&-\mbox{Tr}_I\{V_1\pt{\mathcal{L}_{0I}+{\rm i}\nu}^{-1} V_1
\rho_{SS} \} \label{Snu}
\end{eqnarray}
In Eq.~(\ref{Diff}) we used
$$\alpha=\int_{-1}^1 {\rm d}\cos\theta_0 \cos^2\theta_0 {\cal N}(\cos\theta_0),$$
which gives the angular dispersion of the atom momentum due to the
spontaneous emission of photons. The operator $V_1$ in Eq.~(\ref{Snu}) is given by
\begin{equation}
\label{V1}
V_1=\varphi_L V_L+\varphi_c V_c
\end{equation}
where
\begin{eqnarray}\label{1stlaser}
V_L&=&{\rm i}\hbar\Omega(\sigma\da-\sigma)\\
V_c&=&-\hbar\tilde g(a\sigma\da+a\da \sigma) \label{1stcavity}
\end{eqnarray}
describe respectively the mechanical effects of the drive and of
the cavity at first order in $\eta$, with the two coefficients
\begin{eqnarray}
\varphi_L&=&\cos\theta_L\\
\varphi_c&=&\cos\theta_c\tan\phi
\end{eqnarray}
which depend on the geometry of the setup.
Operator~(\ref{V1}) is the gradient of the atom-field interaction at the center of
the trap and corresponds to the mechanical force in the semiclassical limit~\cite{Nienhuis91}.

\subsection{Rate Equation}

From Eq.~(\ref{externaleq}) one can directly derive the rate
equation for the occupation probability $p_n=\br{n}\mu \ke{n}$ of
the phonon number state $\ke{n}$, namely
\begin{eqnarray}
\frac{d}{dt}p_n&=&\eta^2\left[(n+1)A_-p_{n+1} \right.\nonumber\\
&&\left.-((n+1)A_++nA_-)p_n+nA_+p_{n-1}\right]\label{RateEq:pn}
\end{eqnarray}
where
\begin{eqnarray}
\label{Apm} A_{\pm}=2\mbox{Re}\pg{S(\mp\nu)+D}
\end{eqnarray}
are the rate of heating ($A_+$) and cooling
($A_-$). The solution of this type of equation
is well known~\cite{Stenholm86}. The average phonon number obeys
the equation
\begin{eqnarray}\label{evolution}
\dot{\av{n}} =-\eta^2 (A_{-}-A_{+})\av{n}+\eta^2 A_{+}
\end{eqnarray}
which, for $A_{-}>A_{+}$, has solution
\begin{equation}
\av{n}_t=\av{n}_0{\rm e}^{-Wt}+\av{n}_{\rm St}\left(1-{\rm
e}^{-Wt}\right)
\end{equation}
Here, $\av{n}_0$ is the initial average phonon number and
\begin{eqnarray}
\label{n:steady} \av{n}_{\rm St}=\frac{A_+}{A_- - A_+}
\end{eqnarray}
is the average phonon number at steady state,
while
\begin{eqnarray}
\label{Cool:Rate} W=\eta^2 (A_- - A_+)
\end{eqnarray}
is the cooling rate.

\subsection{Discussion}

In Eq.~(\ref{RateEq:pn}) the internal dynamics enter through the
coefficients $S(\nu)$ and $D$, which determine the
rates~(\ref{Apm}). The function $S(\nu)$ is the spectrum of the
fluctuations of the radiative force on the atom, namely the
Fourier transform of the autocorrelation function of the operator
$V_1$ in Eq.~(\ref{V1}). For the atom coupled to an optical
resonator and driven transversally by a laser, we use the
definition~(\ref{V1}) in Eq.~(\ref{Snu}), and obtain
$$S(\nu)=\varphi_L^2 S_L(\nu)+\varphi_c^2S_c(\nu)+\varphi_L \varphi_c S_{cL}(\nu
)$$ Here,
$$S_L=-\mbox{Tr}_I\{V_L\pt{\mathcal{L}_{0I}+{\rm i}\nu}^{-1} V_L \rho_{\rm St} \}$$
is the contribution of the mechanical effect due to the laser, the
term
$$S_c=-\mbox{Tr}_I\{V_c\pt{\mathcal{L}_{0I}+{\rm i}\nu}^{-1} V_c \rho_{\rm St} \}$$
the contribution of the mechanical effect due to the resonator, and
\begin{eqnarray*}
S_{cL}&=&-\mbox{Tr}_I\{V_c\pt{\mathcal{L}_{0I}+{\rm i}\nu}^{-1} V_L \rho_{\rm St} \}
\\
        & &-\mbox{Tr}_I\{V_L\pt{\mathcal{L}_{0I}+{\rm i}\nu}^{-1} V_c \rho_{\rm St}
\}
\end{eqnarray*}
the contribution due to correlations between the mechanical
effects of laser and resonator. Depending on the geometry of the
setup, one term can be dominant over the others.

The coefficient $D$, Eq.~(\ref{Diff}), gives the diffusion in the dynamics of the
center-of-mass motion. It is the product of two terms: the
spontaneous emission rate of the excited state into the modes of
the e.m.-field, and the stationary excited state population, which is determined by the overall dynamics at zero order in the Lamb-Dicke expansion.

\section{Cavity cooling of trapped atoms}
\label{Sec:III}

\subsection{An explicit form of the rate equation for cooling}

An analytical form for the rates entering
equation~(\ref{RateEq:pn}) can be derived in the limit of a weak
laser drive. The main steps of the derivation are reported in
Appendix A. In this limit the heating and cooling rates take the
form
\begin{eqnarray}\label{app:rateweak}
A_{\pm}
&=&\gamma\alpha|{\cal T}_S|^2+\gamma|\varphi_L{\cal T}_L^{\gamma,\pm}+\varphi_c
{\cal T}_c^{\gamma,\pm}|^2\\
& &+\kappa|\varphi_L{\cal T}_L^{\kappa,\pm}+\varphi_c{\cal T}_c^{\kappa,\pm}|^2
\nonumber
\end{eqnarray}
with
\begin{eqnarray}\label{transitions}
{\cal T}_S&=&\Omega\frac{\delta_c+{\rm i}\kappa/2}{f(0)}\\
{\cal T}_L^{\gamma,\pm }&=&{\rm i}\Omega\frac{(\delta_c\mp\nu+{\rm i}\kappa/2)}
{f(\mp\nu)}\\
{\cal T}_L^{\kappa,\pm }&=&{\rm i}\Omega\frac{\tilde g}{f(\mp\nu)}\\
{\cal T}_c^{\gamma,\pm }&=&-\Omega\frac{\tilde g^2(2\delta_c\mp\nu+{\rm i}\kappa
)}{f(0)f(\mp\nu)}  \\
{\cal T}_c^{\kappa,\pm}&=&-\Omega\frac{\tilde
g\pq{(\Delta\mp\nu+{\rm i}\gamma/2) (\delta_c+{\rm
i}\kappa/2)+\tilde g^2}}{f(0)f(\mp\nu)}\label{transitions:F}
\end{eqnarray}
and
\begin{eqnarray}
f(x)=(x+\delta_c+{\rm i}\kappa/2)(x+\Delta+{\rm i}\gamma/2)-\tilde g^2
\end{eqnarray}
The analytic form of Eqs.~(\ref{app:rateweak})-(\ref{transitions:F}) allows one for a
more transparent reading of these complex dynamics, which can be
mapped back to the processes shown in Fig.~\ref{Fig:3}. The rates
are the incoherent sum of three contributions: The first term,
$\gamma\alpha|{\cal T}_S|^2$, describes a change in the motional
state by mechanical coupling to the modes external to the cavity,
namely by the recoil associated with the spontaneous emission of a
photon. It corresponds to the process depicted in
Fig.~\ref{Fig:3}(a) and determines the diffusion coefficient
through the relation
\begin{eqnarray}
D&=&\gamma\alpha|{\cal T}_S|^2/2.
\end{eqnarray}

The second term, $\gamma|\varphi_L{\cal T}_L^{ \gamma,\pm
}+\varphi_c{\cal T}_c^{\gamma,\pm }|^2$, describes scattering of a
laser photon into the external modes by mechanical coupling to the
laser (${\cal T}_L$) and to the cavity (${\cal T}_c$) field. The
two transition amplitudes correspond to the processes depicted
in Fig.~\ref{Fig:3}(b) and (c), respectively. They add up coherently
and may interfere. Note that these
processes, together with the diffusive process, are dominant for
$\kappa\ll\gamma$.

The third term, $\kappa|\varphi_L{\cal T}_L^{\kappa,\pm
}+\varphi_c{\cal T}_c^{\kappa,\pm}|^2$, describes scattering of a
laser photon into the external modes of the electromagnetic field by cavity decay.
The two amplitudes, appearing in this term, correspond to the
processes depicted in Fig.~\ref{Fig:3}(d) and (e), respectively.
Also in this case they add up coherently
and may interfere. This term is dominant for $\kappa\gg\gamma$.

Equation~(\ref{app:rateweak}) contains the basic features of the
dynamics of cavity cooling of trapped atoms. It has been derived
(i) in the Lamb-Dicke regime, (ii) assuming that the electronic
states are bound by the same center-of-mass potential, (iii) in
the limit in which the laser is a weak perturbation to atom and
cavity dynamics. Moreover, it has been derived for one-dimensional
motion. However, since at second order in $\eta$ the rate
equations for the three directions of oscillation decouple in an
anisotropic trap, it can be generalized to three-dimensional
motion as it holds for any geometry of the setup. Below we show
that this equation reproduces and generalizes results found in
some particular regimes~\cite{Vuletic01,Cirac95}.Moreover,
Eq.~(\ref{app:rateweak}) allows one for identifying new parameter
regimes characterized by novel dynamics that lead to efficient
cooling. Some of these dynamics will be presented in
section~\ref{Sec:4:Interfere}.

\subsection{Cooling in the bad cavity limit}

Cooling in the bad cavity limit, as discussed in~\cite{Cirac95},
is recovered by maximizing the ratio $A_-/A_+$ in the limit in
which spontaneous emission is negligible.
In Eq.~(\ref{app:rateweak}) we set
$\gamma=0$ and take $|\delta_c|\gg\nu$. The equivalence with the cooling and
heating rates reported in~\cite{Cirac95} is evident by using the definitions $\tilde \delta=\tilde
g^2\Delta_c/(\kappa^2/4+\Delta_c^2)$ and $\tilde \gamma=
\tilde g^2\kappa/(\kappa^2/4+\Delta_c^2)$, and imposing $\varphi_c=\varphi_L=1$
(namely, the cavity axes and the laser are parallel to the atomic motion). Below we use this
notation but keep $\varphi_L$ and $\varphi_c$, thereby allowing for a more general
geometry. From Eq.~(\ref{app:rateweak}) we find
\begin{eqnarray}
\label{bad:c}
A_{\pm}&\simeq&\frac{\Omega^2\tilde\gamma}{[(\Delta-\tilde\delta)^2+\tilde\gamma
^2/4]}|a_{\pm}|^2\\
a_{\pm}& =    &\varphi_c\pq{1+\frac{2(\tilde\delta-{\rm
i}\tilde\gamma/2)}{\Delta- \tilde\delta\mp\nu+{\rm
i}\tilde\gamma/2}}\nonumber\\
& & -{\rm i}\varphi_L\frac{\Delta-\tilde\delta+{\rm
i}\tilde\gamma/2}{\Delta-\tilde\delta\mp\nu+ {\rm
i}\tilde\gamma/2}\nonumber
\end{eqnarray}
Here, two processes interfere, namely the process in which the
vibrational number changes by one phonon by absorbing a laser
photon, depicted in Fig.~\ref{Fig:3}(d) and described by the term
${\cal T}_L^{\kappa}$ in Eq.~(\ref{app:rateweak}), and the process in which the
vibrational number changes by one phonon by scattering a cavity
photon, depicted in Fig.~\ref{Fig:3}(e) and described by the term
${\cal T}_c^{\kappa}$ in Eq.~(\ref{app:rateweak}). These dynamics are due
to correlations between the mechanical effects of laser and the cavity, and
they depends critically on the geometric setup, as it is visible
from Eq.~(\ref{bad:c}).

\subsection{Sideband cooling in the good cavity limit}
\label{Sec:4:Bad}

We consider the case where the atom is far-off resonance from the
cavity and the laser, $|\Delta|\gg\gamma,g,\kappa$.
At leading order in $\Delta$ the rates of heating and cooling take
the form
\begin{eqnarray}
A_{\pm}=\frac{\Omega ^2}{\Delta^2}
\left[(\alpha+\varphi_L^2)\gamma +
\frac{\tilde g^2 \kappa (\varphi_L^2 +\varphi_c^2) }{\kappa^2/4 + (\delta_c \mp
\nu )^2} \right]+{\rm O}(1/\Delta^3)\nonumber\\
\label{Apm:Side}
\end{eqnarray}
In this parameter regime there is no relevant contribution to the
mechanical effects from correlations between cavity and laser
dipole force. Cooling is found for $\delta_c<0$, and the
corresponding average phonon number at steady state is
\begin{eqnarray}
\label{n:Sideband}
\av{n}_{\rm St}^{(\Delta)}&=&\frac{\frac{\kappa^2}{4} + (\delta_c + \nu )^2}{4(-\delta_c)
\nu }(1+B)
\end{eqnarray}
with
\begin{eqnarray}
B=\frac{\gamma}{\tilde
g^2\kappa}\left(\frac{\alpha+\varphi_L^2}{\varphi_L^2 +\varphi_c^2}
\right)\left(\frac{\kappa^2}{4} + (\delta_c - \nu )^2\right)
\end{eqnarray}
In the following we do not discuss the solutions leading to Doppler cooling, and focus onto the parameter regimes that lead to ground state cooling, assuming ${\gamma>\nu}$.

For $\kappa\ll\nu$ Eq.~(\ref{n:Sideband}) reaches the minimum value at
$\delta_c=-\nu$,
\begin{eqnarray}
\label{n:min}
\av{n}_{\rm St}^{(\Delta)}\Bigl|_{\delta_c=-\nu}
=\frac{\kappa ^2}{16 \nu ^2} + \frac{1}{4 C_1}\left(\frac{\alpha+\varphi_L^2}{\varphi_L^2 +
\varphi_c^2}\right) \left(1+\frac{\kappa ^2}{16 \nu^2}\right)
\end{eqnarray}
where
\begin{equation}
\label{C1}
C_1=\tilde g^2/\gamma\kappa
\end{equation}
is the one-atom cooperativity~\cite{Kimble94}. The corresponding cooling rate is

\begin{eqnarray}
\label{W:min}
W\Bigl|_{\delta_c=-\nu}
=\eta^2 4C_1(\varphi_L^2 +\varphi_c^2)\frac{\Omega^2}{\Delta^2}\gamma
\left(1-\frac{1}{1+(4\nu/\kappa)^2}\right)
\end{eqnarray}
Therefore, large ground-state populations and large cooling rates can be achieved
for $\nu\gg\kappa$ and $C_1\gg 1$, namely for good cavities and in
the limit in which the cavity linewidth is much smaller than the
trap frequency.

Insight into these results can be found by using the dressed state picture discussed in Sec.~\ref{Sec:II}: For
$\kappa\ll\nu$, large $\Delta$ and $\delta_c=-\nu$ the excitation
spectrum corresponds to the situation depicted in Fig~\ref{Fig:2}(a), where the the red sideband transition is
resonant with the narrow resonance at frequency $\lambda_+=\nu$, while the carrier and the blue
sideband are driven far-off resonance. Hence, this condition is analogous to sideband
cooling, whereby now the narrow resonance is the dressed state of
the system composed by cavity and atom.

The cavity loss rate sets the lower limit to the width of the narrow resonance, on
which sideband cooling is made, and thus to the efficiency of the
process, as it is visible in Eqs.~(\ref{n:min}) and~(\ref{W:min}).
From these equations it is also visible that large cooperativities ensure better efficiencies.

The results reported in this section have been obtained from rate equation~(\ref{RateEq:pn}) with the coefficients Eqs.~(\ref{app:rateweak})-(\ref{transitions:F}) expanded at leading order in $1/\Delta$. Such expansion is valid in the limit where the detuning between atom and cavity
(respectively, laser) is the largest physical parameter. We remark that
Eq.~(\ref{n:min}) is in agreement with the result reported
in~\cite{Vuletic01} in the corresponding parameter regime, whereby the different numerical factors, as
well as the dependence on the angles, are due to the different
laser configuration there considered.

\section{Cooling by quantum correlations in the good cavity limit}
\label{Sec:4:Interfere}

In this section we present and discuss novel cooling dynamics based on quantum interference effects, which are dominant in the good cavity limit,
when $$\kappa\ll\nu,\gamma,\tilde g.$$ In this regime we focus onto the
first two terms of the rates~(\ref{app:rateweak}), corresponding to the processes
in Fig.~\ref{Fig:3}(a)-(c), and treat cavity decay, giving rise to the processes depicted in Fig.~\ref{Fig:3}(d),(e), as small
perturbations.
In the following we assume that $$\phi\neq \frac{\pi}{2}(2n+1)$$ namely the coupling to the cavity
mode does not vanish at zero order in the Lamb-Dicke expansion, and it is given by Eq.~(\ref{tilde:g}).
Comparisons among the efficiencies of the cooling schemes are presented in Sec.~\ref{Sec:Results}.

\subsection{Discussion}

Efficient cooling is achieved by maximizing the rate $A_-$
together with the ratio $A_{-}/A_{+}$. In this way, ideally one
maximizes the cooling rate, Eq.~(\ref{Cool:Rate}), and minimizes
the average number of phonons at steady state,
Eq.~(\ref{n:steady}). In general, by inspection of
Eqs.~(\ref{app:rateweak})-(\ref{transitions:F}) one can identify a
strategy for maximizing $A_-/A_+$, consisting in identifying the
parameters such that the heating and/or diffusion processes
vanish. In this regime, the optimal parameters that maximize $A_-$
are found whenever
\begin{equation}
\label{fx}
\mbox{Re}\{f(\nu)\}=0,
\end{equation} thereby minimizing the denominator
of $A_-$. Physically, this corresponds to set the red sideband
transition at a resonance of the atom-cavity system. This strategy is effective when the linewidth of the corresponding resonance is smaller than the trap frequency. Equation~(\ref{fx})
leads to a condition that relates the
cavity detuning $\delta_c$ with the atom detuning $\Delta$, namely
\begin{eqnarray}\label{Delta}
\Delta_{\rm opt}(\delta_c)\equiv \frac{\tilde
g^2+\gamma\kappa/4}{\delta_c+\nu}-\nu
\end{eqnarray}
where we assume fixed couplings and decay rates, hence also a
fixed cooperativity. For instance, in the case of sideband cooling discussed in Sec.~\ref{Sec:4:Bad},
the optimal cooling conditions are reached for $\delta_c=-\nu$, corresponding
to the solution of Eq.~(\ref{fx}) for $\Delta\to\infty$. In this limit, the linewidth of the dressed state resonance which is used for cooling is infinitely small, and the steady state occupation vanishes accordingly.

Below we discuss various regimes where
ground-state cooling is efficient and which may be identified, for the
corresponding parameter regimes, with an approximate solution of
Eq.~(\ref{fx}).

\subsection{Suppression of diffusion by quantum interference}\label{rateeqk0}

In this section we discuss a cooling scheme based on the suppression of diffusion by quantum interference.
The cooling dynamics  are based on the suppression of the carrier transition, and can be understood with the dressed state picture. As discussed in Sec.~\ref{Sec:II}, the carrier transition can vanish in the regime in which laser and cavity are resonant. The sidebands due to the harmonic motion, however, are weak perturbations in opposition of phase with respect to the carrier. Thus, they give rise to photon scattering with probability given by the corresponding value of the excitation spectrum, depicted in Fig.~\ref{Fig:2}(b) for some parameter regime. The cooling strategy is thus to enhance the red sideband over the blue sideband absorption, thereby profiting of the suppression of carrier excitations, and thus of diffusion.
This idea is reminiscent of cooling mechanisms based on quantum interference between atomic levels~\cite{Morigi03,Morigi00}, whereby here the suppression of the carrier transition is due to the destructive interference between the laser and the light elastically scattered into the resonator by the atom.

Diffusion is suppressed when ${{\cal T}_S=0}$
in Eqs.~(\ref{app:rateweak}), leading to the vanishing of the diffusion coefficient $D$, Eq.~(\ref{Diff}).
From Eq.~(\ref{app:rateweak})
this occurs when $\delta_c=0$ and, ideally, for $\kappa=0$. Let us first consider the ideal condition of a lossless resonator. In this case, for $\delta_c=0$ the steady state average phonon number is given by
\begin{eqnarray}
\av{n}_{\rm St}^{(0)}=\frac{[\nu(\nu+\Delta)-\tilde{g}^2]^2
+\gamma^2\nu^2/4}{4\nu\Delta(\tilde{g}^2-\nu^2)}
\end{eqnarray}
Cooling is achieved when either the
relations $\Delta>0$ and $\tilde{g}>\nu$, or the relations
$\Delta<0$ and $\tilde{g}<\nu$, are fulfilled.
The minimum for $\av{n}_{\rm St}^{(0)}$ is obtained when
$\Delta=\Delta_{\rm opt}(\delta_c=0)$, see Eq.~(\ref{Delta}).
For these values the minimum number of phonon at steady state is
\begin{eqnarray}
\label{n:CIT}
\av{n}_{\rm St}^{(0)}\Bigl|_{\Delta_{\rm opt}(0)}
&=&\frac{\gamma^2}{16\Delta_{\rm opt}(0)^2}\nonumber\\
&=&\frac{\gamma^2\nu^2}{16(\tilde g^2-\nu^2)^2}
\end{eqnarray}
with the corresponding cooling rate
\begin{eqnarray}
\label{W:CIT} W\Bigl|_{\Delta_{\rm opt}(0)}=4\eta^2
(\varphi_L^2
+\varphi_c^2)\frac{\Omega^2}{\gamma}\left(1-\frac{1}{1+(4\Delta_{\rm opt}(0)/\gamma)^2}
\right)\nonumber\\
\end{eqnarray}
Therefore, ground state cooling, namely $\av{n}_{\rm St}^{(0)}\ll 1$,
is found when $\Delta=\Delta_{\rm opt}$ and $|\Delta|\gg\gamma$, or
equivalently for $|\tilde{g}^2/\nu-\nu|\gg\gamma$. This condition can be
fulfilled (i) when $\nu\gg\gamma$ and
(ii) when $\tilde{g}^2\gg\gamma\nu$, which is most interesting as it can give ground state cooling even when $\nu\ll\gamma$. Below we discuss these two cases in detail.

Case (i) corresponds to the so-called strong confinement
regime~\cite{Eschner03}, namely when the linewidth of the dipole
transition is smaller than the trap frequency. In this case
sideband cooling is efficient in free space (i.e., in absence of the resonator). Like for sideband
cooling in free space, the implementation inside a cavity
resonator requires $\Delta= -\nu$, leading to the final occupation $\av{n}_{\rm St}^{(0)}\approx\gamma^2/16\nu^2$.

Case (ii) can be fulfilled in the so-called weak confinement
regime~\cite{Eschner03}, namely when the linewidth of the dipole
transition is much larger than the trap frequency. Result~(\ref{n:CIT}) shows that ground state cooling can be efficiently achieved.
This is to our knowledge a novel regime. Here, $\Delta\sim \tilde{g}^2/\nu$, so
that we can rewrite the cooling limit as
\begin{equation}
\label{n0:CIT:ideal}
\av{n}_{\rm St}^{(0)}\Bigl|_{\Delta_{\rm opt}(0)}\approx
\gamma^2\nu^2/16 \tilde{g}^4.
\end{equation}

We now discuss how the efficiency is modified for $\delta_c=0$ but $\kappa$ finite. At first order in $\kappa$ the diffusion coefficient $D=0$. In fact, $D={\rm O}(\kappa^2)$, being the stationary population of the excited state of second order in $\kappa$ in this regime~\cite{Zippilli04a,Zippilli04b}.
At first order in this expansion the steady state average phonon number is
\begin{eqnarray}
\label{n:K}
\av{n}_{\rm St}=\av{n}_{\rm St}^{(0)}(1+F)
\end{eqnarray}
where the term
\begin{eqnarray}
F&=&\frac{\kappa ^2}{\nu ^2}C_1 \frac{ \gamma }{2}{\cal A}_-  -
2\frac{\kappa}{\nu}\frac{{\cal A}_-}{{\cal A}_--{\cal A}_+}\frac{\varphi_L
\varphi_c
 }{\varphi_L^2+\varphi_c^2}
\end{eqnarray}
is the correction at first order in $\kappa$, $C_1$ is the one-atom cooperativity defined in Eq.~(\ref{C1}),
and
\begin{eqnarray}
{\cal A}_\pm
=\frac{ \nu^2\gamma}{[\nu(\nu\mp\Delta)-\tilde g^2]^2+\nu^2\gamma^2/4}
\label{apm0}
\end{eqnarray}
Cavity decay increases the linewidth of the dressed-state resonances, and it
thus detrimental. Nevertheless, for high cooperativities the result
we find in Eq.~(\ref{n:K}) approaches the result of the
ideal case, Eq.~(\ref{n0:CIT:ideal}). In particular, for $\tilde g\gg\nu$ its value at $\Delta=\Delta_{\rm opt}(0)$ takes the simple form
\begin{equation}
\label{n:CIT:C1} \av{n}_{\rm St}\Bigl|_{\Delta_{\rm opt}(0)}\approx\av{n}_{\rm St}^{(0)}\Bigl|_{\Delta_{\rm opt}(0)}+\frac{1}{8C_1}
\end{equation}
showing that the corrections scale with $1/C_1$.

It must be remarked that the equations presented in this section have
been obtained from Eqs.~(\ref{app:rateweak}) in the limit of weak
coupling. Nevertheless, they are also valid when the dipole is driven by a saturating laser field. In that case, at zero order in the Lamb-Dicke expansion the atom is in the ground state and the cavity in a coherent state with amplitude
$\beta_c=-\Omega/\tilde g$, such that
the steady state at zero order is~\cite{Alsing92}
\begin{eqnarray}
\label{Steady:smallk} \rho_{\rm 0St}&=&\kb{g,\beta_c}{g,\beta_c}
\end{eqnarray}
The derivation of the rate equation of cooling, obtained by making {\it no} assumption regarding the
strength of the laser intensity, is reported in appendix~\ref{app:smallk}. The result agrees with the results reported in this section, which have been evaluated from Eqs.~(\ref{app:rateweak})
under the assumption of weak laser fields.
This agreement is not a casuality: In fact, when the conditions for this interference
effect are fulfilled, the atom is driven well below saturation
even for strong laser and cavity fields, since they mutually cancel at the atomic
position~\cite{Zippilli04a,Zippilli04b}. Nevertheless, the sideband transitions take place since they are out of phase with respect to the carrier.

\subsection{Suppression of heating by quantum interference}
\label{Sec:Heating}

In this section we discuss a cooling scheme based on the suppression of heating processes by quantum interference. This interference phenomenon is found in the good cavity limit, and corresponds to the situation in which the heating processes depicted in Fig.~\ref{Fig:3}(b) and~(c) cancel out. The corresponding parameters are identified in
Eq.~(\ref{app:rateweak}) by imposing the condition ${\varphi_L{\cal T}_L^{\gamma,+}+\varphi_c{\cal T}_c^{\gamma,+}=}0$. This condition can be fulfilled
for instance when $\varphi_L=0$, namely when the
laser is orthogonal to the motional axis and therefore exerts no force,
and ${\cal T}_c^{\gamma,+}=0$, namely when the transitions
to the blue sideband induced by the mechanical effects of the resonator vanish.
Below we discuss this particular case.

We assume $\varphi_L=0$. Condition ${\cal T}_c^{\gamma,+}=0$ is fulfilled when
$\delta_c=\nu/2$ and, ideally, $\kappa=0$. In this limit $A_+=\gamma\alpha|{\cal T}_S|^2$
and the average phonon number at steady state is
\begin{eqnarray}
\av{n}_{\rm St}^{(0)}=\alpha\frac{9\gamma ^2 \nu ^2/16+ \left[\tilde g^2
-3\nu  (\Delta +\nu )/2\right]^2}{16\tilde g^4 \varphi_c^2}
\end{eqnarray}
It reaches a minimum for $\Delta=\Delta_{\rm opt}(\nu/2)$, namely $\Delta_{\rm opt}(\nu/2)=2\tilde g^2/3\nu-\nu$, that has the form
\begin{eqnarray}
\label{n:0:heat}
\av{n}_{\rm St}^{(0)}\Bigl|_{\Delta_{\rm opt}(\nu/2)}=\frac{9\alpha}{16\varphi_c^2}
\frac{\gamma^2\nu^2}{16\tilde g^4}
\end{eqnarray}
with the corresponding cooling rate
\begin{eqnarray}
\label{W:0:heat} W\Bigl|_{\Delta_{\rm opt}(\nu/2)} &=&16\eta^2
\frac{\Omega^2}{\gamma}\frac{\varphi_c^2}{(1+3\nu^2 /4\tilde
g^2)^2+(3\gamma \nu /8\tilde g^2)^2} \nonumber\\
\end{eqnarray}
To our knowledge, this is a novel cooling regime.
Insight into these dynamics cannot be simply gained by inspection
of the excitation rate at zero order in the Lamb-Dicke parameter.
In fact, the disappearance of the blue sideband transition is due
to a quantum interference effect between the paths of mechanical
excitation driven by the resonator. Comparing this case with cooling
by suppression of diffusion, see Sec.~\ref{rateeqk0}, one finds that for $\delta_c=\nu/2$ one can
reach lower temperatures in a faster time, as it is evident from a
comparison of Eqs.~(\ref{n:0:heat}) and~(\ref{W:0:heat})
with Eqs.~(\ref{W:CIT}),~(\ref{n:CIT:C1}) and with the results reported in Sec.~\ref{Sec:4:Bad}.

A finite, but small, value of $\kappa$ leads to the corrected average excitation
\begin{eqnarray}
\av{n}_{\rm St}=\av{n}_{\rm St}^{(0)}\pt{1+\kappa
F}+\kappa G
\end{eqnarray}
where
\begin{eqnarray}
F&=&\frac{1}{2}  \left[\frac{\tilde g^2 \gamma  }{9 \gamma ^2 \nu ^2
/16+\left(\tilde g^2 -3\nu  (\Delta +\nu   )/2\right)^2}\right.
\nonumber\\
&&\left.+\frac{\left(\gamma ^2/4+(\Delta +\nu )^2\right) }{\tilde g^2
\gamma }-\frac{2 (\Delta +\nu )}{\gamma  \nu }\right]\\
G&=&\frac{9 \gamma ^2 \nu ^2/16+\left[\tilde g^2 -3 \nu
 (\Delta +\nu )/2\right]^2 }{4 \tilde g^2 \gamma  \nu ^2}
\end{eqnarray}
are the corrections at first order in $\kappa$. They lower the efficiency of the mechanism. In particular,
the optimal final occupation number becomes
\begin{eqnarray}
\av{n}_{\rm St}\Bigl|_{\Delta_{\rm opt}(\nu/2)}
                        &=&\left(1+\frac{1}{8 C_1}\right)\av{n}_{\rm St}^{(0)}\Bigl|_{\Delta_{\rm opt}(\nu/2)} \nonumber\\
& &+ \frac{\alpha/ \varphi_c^2
+9}{64 C_1}
\end{eqnarray}
while the corrections to the cooling rate scale with $1/C_1$. Therefore, for large cooperativities this interference effect is relevant to the cooling dynamics. We remark, that as for $\delta_c=\nu/2$ the heating transition vanishes, similarly for $\delta_c=-\nu/2$ the cooling transition cancels out.

\subsection{Suppression of diffusion and heating by quantum interference}
\label{standingwavedrive}

We finally discuss a cooling scheme based on the suppression of both diffusion and heating transitions by quantum interference. Let us first consider suppression of the carrier excitation, that leads to a vanishing diffusion coefficient. This can be achieved by using a standing-wave drive, such that the trap center is at one of its nodes. Therefore, at zero order in the Lamb-Dicke expansion the atom does not scatter any photon
and the cavity is thus empty. Photon scattering originates from the dynamics due to the
finite size of the atomic wave packet, and it is thus a process of second
order in the Lamb-Dicke expansion. In order to investigate these
dynamics, we evaluate the heating and cooling rate entering the rate equation~(\ref{RateEq:pn})
by considering a new Hamiltonian, which is given by operator~(\ref{H:tot}) with the new
laser-dipole coupling
\begin{eqnarray}\label{HLstandingwave}
H_L=\hbar\Omega\cos(kx\cos\theta_L+\phi_L)(\sigma\da+\sigma)
\end{eqnarray}
The condition for which the trap is at a node of the laser standing wave corresponds to choose
$$\phi_L=\pi/2,$$
For this value, the interaction with the laser vanishes at zero order in the Lamb-Dicke expansion and the steady state is the empty cavity field and the atom in the ground state, namely
$\rho_{\rm St}'=\kb{g,0_{c}}{g,0_{c}}$. Note that no assumption has been made on the value of the Rabi frequency $\Omega$. By expanding Eq.~(\ref{HLstandingwave}) at first order in the Lamb-Dicke parameter, we obtain in place of the operator~(\ref{V1}) the new interaction term
$$V_{L}=-\hbar\Omega\varphi_L(\sigma\da+\sigma)$$
In the rest of this section we will consider  $\varphi_L=1$.

Following the lines of the derivation as in section~\ref{rateeq}
with the new definitions, we obtain the equation for the external
dynamics, Eq.~({\ref{externaleq}}), where now
$$D=0$$
and
\begin{eqnarray}\label{S'}
S(\nu)=-\mbox{Tr}_I\{V_{L}\pt{\mathcal{L}'_{0I}+i\nu}^{-1} V_{L} \rho_{St}' \}
\end{eqnarray}
Here, $\mathcal{L}'_{0I}$ has the same form as $\mathcal{L}_{0I}$ in
Eq.~(\ref{Liouv0I}), with $H_{0L}=0$.
Clearly, the disappearance of the diffusion term is due to the fact that
there is no field at zero order in the Lamb-Dicke expansion.

The term~(\ref{S'}), giving the mechanical action on the atomic
motion, originates solely from scattering of laser photons. In
fact, the mechanical effects of the resonator field appear at
higher order in the Lamb-Dicke expansion. The cooling and heating
rates $A_\pm'=2 \mbox{Re}\{S(\mp\nu)\}$ take the form
\begin{eqnarray}
\label{Apm'}
A'_\pm=\gamma|{\cal T}_{1L}^{\gamma\pm }|^2+\kappa|{\cal T}_{1L}^{\kappa\pm}|^2
\end{eqnarray}
where now
\begin{eqnarray}
{\cal T}_{1L}^{\gamma,\pm }&=&\Omega\frac{(\delta_c\mp\nu+{\rm i}\kappa/2)}
{f(\mp\nu)}\\
{\cal T}_{1L}^{\kappa,\pm }&=&\Omega\frac{\tilde g}{f(\mp\nu)}\\
\end{eqnarray}
and
\begin{eqnarray}
f(x)=(x+\delta_c+{\rm i}\kappa/2)(x+\Delta+{\rm i}\gamma/2)-\tilde g^2
\end{eqnarray}
Hence, the transition amplitudes do not relevantly differ from the ones in
Eqs.~(\ref{transitions})-(\ref{transitions:F}). However, no low saturation limit
is needed in the derivation of these results.

We study now the cooling dynamics in an exemplary limit, namely in the case of
a very good resonator. We first consider an ideal resonator, namely $\kappa=0$. We obtain
\begin{eqnarray}
A'_\pm|_{\kappa=0} =\frac{\Omega^2 {\left(\delta_c \mp
\nu  \right) }^2\gamma}{\gamma^2 \,\left( \delta \mp \nu
\right)^2/4 +{\left[  \left( \delta  \mp \nu  \right) \,\left(
\Delta \mp \nu  \right) -\tilde g^2 \right] }^2}
\end{eqnarray}
Thus, when $\delta_c=\nu$ then the heating transition vanishes. Since the diffusion is also zero, then $A'_+=0$ and
$$\av{n}_{\rm St}^{(0)}= {\rm o}(\eta^2).$$
The corresponding cooling rate reaches the maximum value for $\Delta=\Delta_{\rm opt}(\nu)$, namely $\Delta_{\rm opt}(\nu)=(\tilde g^2-2\nu^2)/2\nu$, and takes the form
$$W=4\eta^2\frac{\Omega^2}{\gamma}.$$
This result is exact for $\kappa=0$.

Finite values of $\kappa$ introduce corrections to the heating rate, which takes the form
$$A'_+\simeq \kappa\Omega^2/\tilde g^2=\frac{\Omega^2}{\gamma}
\frac{1}{C_1}$$
Correspondingly, the average number of phonon
at steady state is
\begin{eqnarray}
\av{n}_{\rm St}\simeq\frac{[2\nu(\Delta+\nu)-\tilde
g^2]^2+\gamma^2\nu^2}{4 \nu^2 \gamma^2}\frac{1}{C_1}
\end{eqnarray}
 the minimum value of the number of excitations at steady state
is found at $\Delta=\Delta_{\rm opt}(\nu)$ and is given by
$$\av{n}_{\rm St}\Bigl|_{\Delta_{\rm opt}(\nu)}= 1/4C_1,$$ These
cooling dynamics are novel, and correspond to the situation in
which the excitation pathways of the combined multilevel
atom--cavity system interfere destructively, thereby suppressing
the blue sideband excitation. They are reminiscent of cooling
schemes for multilevel atoms discussed in~\cite{Evers}, where
suppression of the carrier and blue sideband transitions is
achieved by quantum interference between atomic excitations. In
the case studied here, however, the mechanism which leads to
suppression of the carrier transition is different from the one
that leads to the suppression of the blue sideband transition, and
both are due to the composite effect of cavity and laser on the
atom. Moreover, the parameter regime here considered is one of
several possible, that can be identified by imposing the
disappearance of the blue sidebands transition.

\section{Results}
\label{Sec:Results}

In this section we compare the cooling efficiencies in the various
regime, as evaluated from the analytical results, and check the
range of validity of the analytical calculations with a quantum
Monte-Carlo wavefunction method, where the full quantum dynamics
of master equation~(\ref{meq}) is simulated. We focus onto the
good cavity limit, in particular onto the parameters
$\kappa\ll\nu\ll\gamma$.

\subsection{Plot of the analytical results}

\begin{figure*}[t]
\includegraphics[width=18cm]{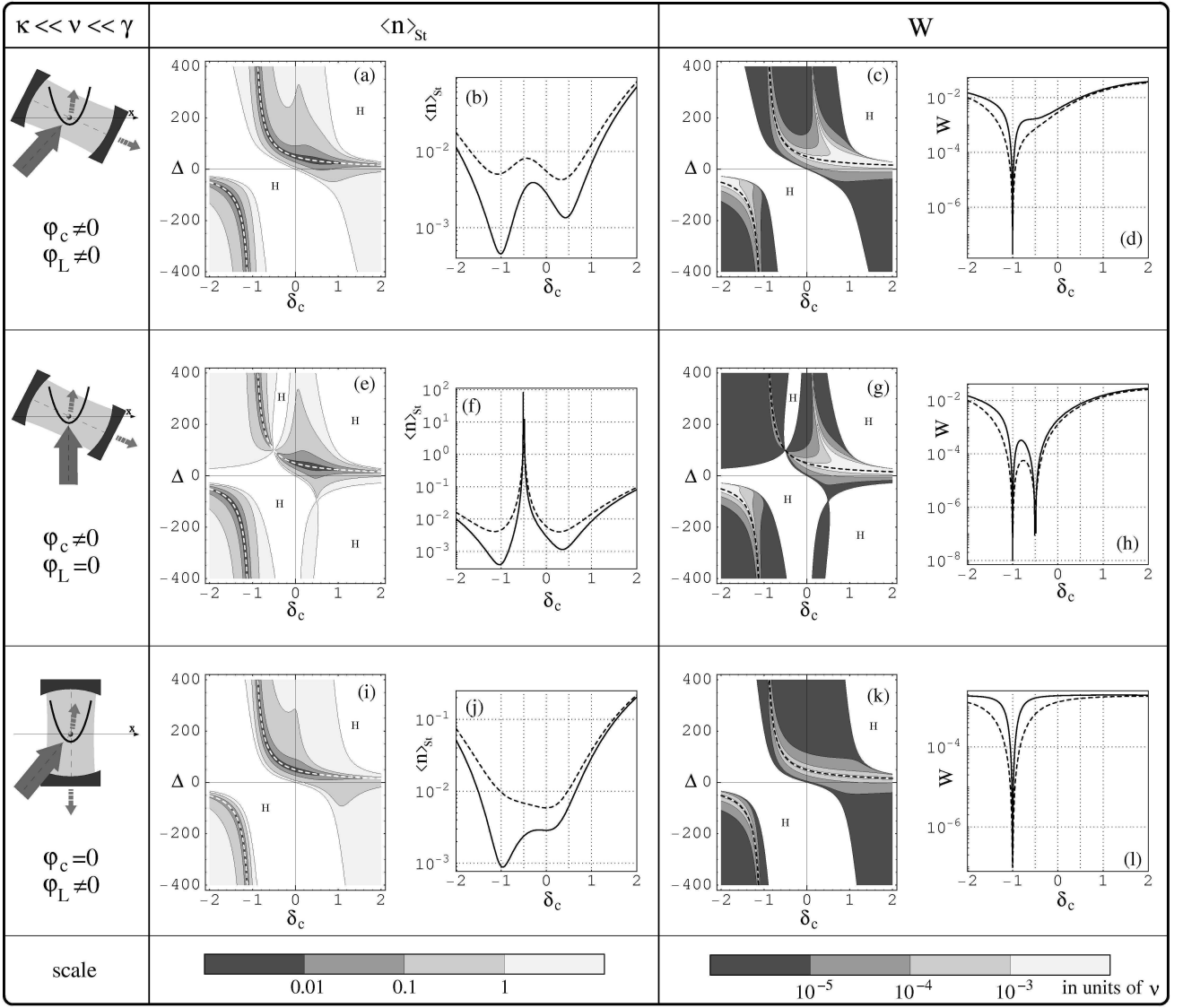}
\caption{Average phonon number at steady state $\av{n}_{\rm St}$
and corresponding cooling rate $W$, in units of $\nu$, in the good
cavity limit, for $\kappa\ll\nu\ll\gamma$ and for three possible
geometries: in (a)-(d) (first row) both cavity and laser fields
contribute to the cooling: Here $\theta_L=\theta_c=\pi/4$. In
(e)-(h) (second row) the mechanical effects of the cavity solely
determine cooling: Here $\theta_L=\pi/2$ and $\theta_c=\pi/4$. In
(i)-(l) (last row) the mechanical effects of the laser solely
determine cooling: Here $\theta_L=\pi/4$ and $\theta_c=\pi/2$. The
contour plots show $\av{n}_{\rm St}$ and~$W$ as a function of
$\delta_c$ and $\Delta$ in units of $\nu$. The gradation of grey
follows the scale where darkest region corresponds to the smallest
values, the lightest region to the largest values. The
corresponding values are reported at the bottom of the figure. The
heating regions are not coded and explicitly indicated by the
label H. The dashed curve appearing in each contour plot
represents the curve $\Delta_{\rm opt}(\delta_c)$,
Eq.~(\ref{Delta}). The parameters are $\eta=0.1$, $\theta_L=
\pi/4$, $\Omega=\nu$, $\tilde g=7\nu$, $\gamma=10\nu$,
$\kappa=0.01\nu$. The other plots display $\langle n\rangle_{\rm
St}$ and $W$ as a function of $\delta_c$ and $\Delta_{\rm
opt}(\delta_c)$, for the same parameters as of the contour plot
and $\kappa=0.01\nu$ (solid line), $\kappa=0.1\nu$ (dashed line).}
\label{fig:Good}
\end{figure*}

The plots in Figure~\ref{fig:Good} show the phonon number at
steady state and the cooling rate for different geometries of the
setup. In particular, the plots of the first row depict the
situation in which the mechanical effects on the atom are due to
both laser and cavity field, the plots of the second row show the
dynamics when the effects are solely due to the resonator, and the
ones in the third row when the effects are solely due to the
laser. The contour plots show most evidently the parameter regions
where cooling is effective. Here, the dashed line represents the
function~(\ref{Delta}) which determines the parameters minimizing
the steady state temperature. Clearly, in the neighbourhood of
this line the lowest temperature is achieved in all three cases.
We note that the parameter regimes where cooling occurs may differ
depending on whether the dipole forces are due to the resonator or
to the laser.

We now discuss the dynamics in detail. Due to the wealth of
behaviours, we focus onto the parameter regions where ground-state
cooling appears efficient.

Figures~\ref{fig:Good}(b),(f), and~(j) display the value of the
average phonon number as a function of $\delta_c$ and $\Delta_{\rm
opt}(\delta_c)$, namely its value along the
function~(\ref{Delta}). Figures~\ref{fig:Good}(d),(h), and~(l)
show the corresponding cooling rates. Each plot displays two
curves, which have been evaluated for two different values of the
cavity decay rate $\kappa$ (solid curve: $\kappa=0.01\nu$, dashed
curve: $\kappa=0.1\nu$). From these curves it is visible that, as
the cavity decay rate increases, the cooling efficiency decreases,
namely the temperature gets higher and the cooling rate lower.
Nevertheless, for the parameter here considered the cooling
dynamics remains efficient. Let us now discuss the behaviour as we
vary $\delta_c$ and keep $\Delta=\Delta_{\rm opt}(\delta_c)$.

In all cases the function $\av{n}_{\rm St}$ exhibits a minimum at
$\delta_c=-\nu$ at very large values of
$\Delta$. This is the sideband cooling regime, discussed in
Sec.~\ref{Sec:4:Bad}. The cooling rate at these points is very
small, since it scales as $\Delta^{-2}$, as visible from
Eq.~(\ref{W:min})~\cite{Footnote}. This cooling scheme exploits the dressed
states of the system at zero order in the Lamb-Dicke expansion,
see Sec.~\ref{Sec:II}, and its efficiency is thus
relatively independent of whether the cavity or the laser forces
determine cooling.  Here, the cooling efficiency is very sensitive
to variation of $\delta_c$, as visible from the contour plots.
This sensitiveness is due to the narrowness of the linewidth of
the dressed-state resonance which is used for cooling the motion.

The curves in Figures~\ref{fig:Good}(b),(f) show also a minimum at
$\delta_c=\nu/2$. This is due to cooling by suppression of the
resonator's mechanical coupling to the blue sideband transition,
see Sec.~\ref{Sec:Heating}. Clearly, this minimum is more enhanced
in Fig.~\ref{fig:Good}(f), where cooling is solely due to the
mechanical effects of the cavity, and does not appear in
Fig.~\ref{fig:Good}(j), where the resonator field has no
mechanical effects on the atom. The corresponding cooling rate is
relatively large, being the atom driven close to resonance in this
regime. An interesting characteristic, emerging from the contour
plots, is that these cooling dynamics are relatively robust to
fluctuations of the parameters, showing that ground-state cooling
is efficient even in the regime in which suppression of the
heating transition is partial.

The heating region at $\delta_c=-\nu/2$, appearing in the case in
which the mechanical effects of the cavity solely contribute to
cooling, Figs.~\ref{fig:Good}(e)-(h), is originated from the same
interference effects that give rise to cooling at the value
$\delta_c=\nu/2$, and that for $\delta_c=-\nu/2$ leads to
suppression of the cooling transition, see Sec.~\ref{Sec:Heating}.

Another regime where cooling is effective is found at $\delta_c=0$
when the mechanical effects are solely due to the laser. This
parameter regime is characterized by small temperatures and large
cooling rates, as visible from Figs.~\ref{fig:Good}(i)-(l). This
is the regime in which the carrier transition is suppressed by an
interference phenomenon at zero order in the motion, see
Secs.~\ref{Sec:II} and~\ref{rateeqk0}. Cooling efficiency is
robust against fluctuations of the parameters, and appear to be
relatively stable as the value of $\kappa$ is increased, as
compared with the sideband cooling case, see
Fig.~\ref{fig:Good}(j).

In general, we can conclude that in the range of values of
$\delta_c$ around the interval $[0,\nu/2]$, and close to atomic resonance, the
cooling efficiency is relatively high. We remark that the final
temperature is limited by the ratio $\kappa/\nu$. This is
understood in the dressed state picture, as the final limit to the
narrow dressed state resonance is set by the cavity decay rate
$\kappa$.

\subsection{Numerical simulations}

\begin{figure}[h]
\includegraphics[width=8cm]{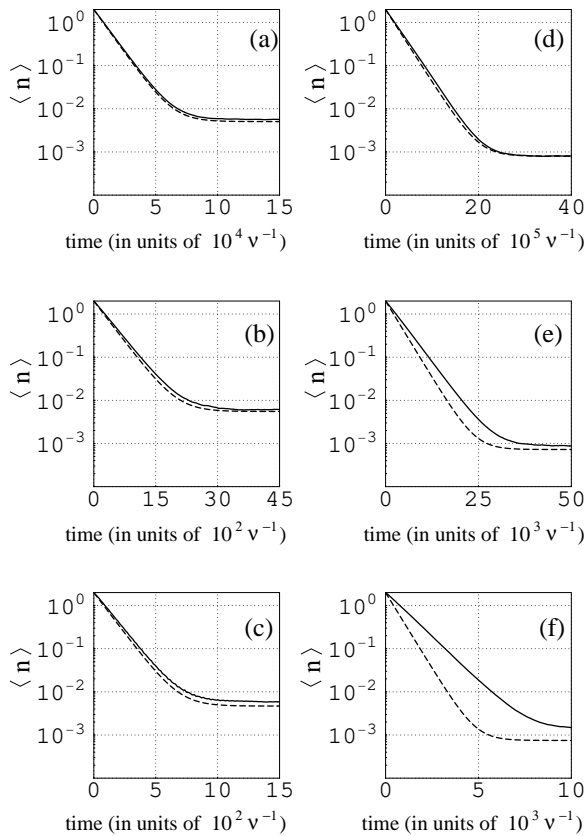}
\caption{Comparison between the analytical
equations, Eqs.~(\ref{RateEq:pn}) and~(\ref{app:rateweak}),
and a full quantum Monte Carlo simulation of Eq.~(\ref{meq}).
The curves show the evolution of the average phonon number as the function of
time in units of $\nu^{-1}$, the dashed lines correspond to the analytical predictions, the solid lines to the
quantum Monte Carlo simulation. The parameters are $\eta=0.1$, $\theta_L=\theta_c=\phi=\pi/4$, $\Omega=\nu$, $\gamma=10\nu$, $\kappa=0.1\nu$ and (a) $g=10\nu$, $\delta_c=-1.1\nu$; (b)
$g=10\nu$, $\delta_c=0$; (c) $g=10\nu$, $\delta_c=0.5\nu$; (d)
$g=50\nu$, $\delta_c=-1.1\nu$; (e) $g=50\nu$, $\delta_c=0$; (f)
$g=50\nu$, $\delta_c=0.5\nu$; } \label{Fig:MC}
\end{figure}

The curves reported in Fig.~\ref{fig:Good} have been obtained from
the analytical equations, which have been evaluated assuming that
dipole and cavity dynamics reach the steady state
on a much faster time scale than the center-of-mass dynamics. In
Fig.~\ref{Fig:MC} we verify these results by comparing them with a
full quantum Monte Carlo wave function simulation of Master
Equation~(\ref{meq}). We see that in general the analytical
predictions are in agreement with the numerical results for a vast
range of parameters, which are experimentally accessible. The
discrepancies are small and are due to parameter regimes where the
adiabatic evolution is not fulfilled. The discrepancies affect
mostly the cooling rate, which varies by an overall factor, while
the final average excitations of the center-of-mass oscillator are
in agreement.

\section{Conclusions}
\label{Sec:Conclusions}

We have presented an extensive study of the cooling dynamics of
trapped atoms in optical resonators.
Our study is based on a rate equation, which we derive from the master equation of the system composed by atom and cavity, and whose validity is
supported by numerical simulations taking into account the full
quantum dynamics. Our analytical results are valid in the Lamb-Dicke regime and when
the center-of-mass potential is independent of the internal state,
like in~\cite{Keller04,Mundt02,KimbleFORT03}.

The equations we derive reproduce the results reported
in~\cite{Vuletic01,Cirac95} in their specific parameter regimes.
Moreover, they allow us to identify new parameter regimes where
the dynamics of the atomic center-of-mass, coupled to a laser and
a cavity field, results from a non-trivial competition of the
laser and of the resonator dipole forces, which can mutually
interfere. These interference effects are at the basis of novel
cooling schemes, which we identify in this paper and which allow
to reach very large ground-state occupations. The corresponding
dynamics are reminescent of cooling schemes exploiting
interference in multi-level atomic
transitions~\cite{Marzoli94,Morigi00,Evers}.

It must be remarked that the interference effects discussed in this work base themselves on the discreteness of the spectrum of the mechanical excitations,
which is the same for the dipolar ground and the excited states.
Dynamics will be substantially modified
when the external potential depends on the internal state, as the
spectroscopic properties of the atom are changed and with it the
scattering cross section. Preliminary considerations are
in~\cite{Morigi03} for the case of trapped multilevel atoms, while
the effects of state-dependent mechanical potentials
on the cavity field have been discussed in~\cite{Rice04}.

In this paper we have focussed onto the case of dipole
transitions, which in free space do not allow one for applying
sideband cooling. Here, ground-state cooling can be obtained in
good resonators, whereby the final cooling efficiency is limited
by the cavity decay rate. The bad-cavity case is contained in the
analytical equations we have derived.  In this limit the effect of
the laser seems to be predominant and for a general configuration
the optimal cooling corresponds to standard cavity sideband
cooling~\cite{Zippilli05b}. This differs strikingly with the good
cavity limit, where correlations between atom and resonator can
lead to very efficient cooling. We remark that these dynamics are
largely modified when the drive is set on the resonator.

To conclude, this work considers the cooling dynamics to the
potential ground state of a single atom inside a resonator. In the
future we will investigate how the collective dynamics of several
atoms, confined in a resonator, influence the cooling efficiency,
and how these can be used to prepare quantum states of the system
in a controlled way.

\acknowledgements

The authors wish to thank Helmut Ritsch and Axel Kuhn for several
stimulating discussions. Support from the IST-network QGATES and of
the Spanish Ministerio de Educaci\'on y Ciencia (Ramon-y-Cajal
fellowship 129170) is acknowledged.

\begin{appendix}

\section{Limit of weak driving field}\label{app:weak}

We consider the limit when the atom is weakly driven, namely when
the Rabi frequency $\Omega$ is much smaller than all the other
physical parameters that characterize the internal dynamics. We
study the dynamics of the motion in perturbation theory in second
order in $\Omega$ and $\eta$, and neglect the terms of order
$\eta^4\Omega^2$, $\eta^2\Omega^4$ and higher.

At zero order in $\Omega$  the laser field is zero. Hence, at
steady state the cavity is empty and the atom is in the ground
state. The state of dipole and cavity is described by the density
matrix $$\rho_{\rm St}^\circ=\kb{g,0_{c}}{g,0_{c}},$$ which is the
solution of the equation $\mathcal{L}_{0I}^\circ\rho_{\rm St}^\circ=0$, with
\begin{eqnarray}
\mathcal{L}_{0I}^\circ\rho&=&\mathcal{L}_{0I}\Bigl|_{\Omega=0}\rho\\
&=&\frac{1}{{\rm i}\hbar}[H_{\rm at}+H_{\rm cav}+H_{0{\rm at-cav}},\rho]+{\cal K}\rho+{\cal L}_{0s}\rho\nonumber
\end{eqnarray}
and is obtained from $\mathcal{L}$, Eq.~(\ref{meq}), at
zero-order in $\eta$ and $\Omega$. Following the general
procedure described in section \ref{rateeq}, we define the
projector
$P\rho=\kb{g,0_{c}}{g,0_{c}}\otimes\sum_n|n\rangle\langle
n|\mbox{Tr}_I\{\langle n|\rho|n\rangle\}$ into the eigenspace of the superoperator
\begin{eqnarray}
{\cal L}_{0}^{\circ}={\cal L}_{0I}^{\circ}+{\cal L}_{0E}
\end{eqnarray}
namely the Liouvillian at zero order in $\eta$ and $\Omega$, at the eigenvalue zero. The effect of the perturbation in $\eta$ and $\Omega$ is described by the equation
$$P\dot\rho={\cal L}_P P\rho$$ where we have eliminated the coupling with other subspaces, thereby obtaining
\begin{eqnarray}
{\cal L}_{\cal P}&=&\sum_{n,m=0}^\infty \eta^{n}\frac{\Omega^{m}}{\tilde g^{m}}
{\cal L}_{n,m}\\
&=& \eta^{2}\frac{\Omega^{2}}{\tilde g^{2}}{\cal
L}_{2,2}+{\rm O}\left(\eta^{4}\frac{\Omega^{2}}{\tilde g^{2}}\right)+{\rm O}\left(\eta^{2}\frac{\Omega^{4}}{\tilde g^{4}}\right)
\end{eqnarray}
At the lowest non-vanishing order only the term ${\cal L}_{2,2}$
is relevant, and
\begin{eqnarray}
{\cal L}_{\cal P}&\simeq&\eta^{2}\frac{\Omega^{2}}{\tilde g^{2}}{\cal L}_{2,2}\\
                 &=     &\varphi_L^2{\cal
L}_L+\varphi_c^2{\cal L}_c+\varphi_c\varphi_L{\cal L}_{cL}+{\cal L}_{\rm diff}
\end{eqnarray}
The subscripts $L$, $c$ and $cL$ label the terms describing
processes in which the mechanical effect on the atoms
are due respectively to the laser, the cavity, and to the cooperative action of laser
and cavity. They have the form
\begin{eqnarray}
{\cal L}_{L}&=&-P{\cal L}_{1L}{\cal L}_{0}^{\circ -1}{\cal L}_{1L}\\
{\cal L}_{c}&=&-P{\cal L}_{1c}{\cal L}_{0}^{\circ -1}
{\cal L}_{1c}{\cal L}_{0}^{\circ -1}{\cal L}_{0L}
{\cal L}_{0}^{\circ -1}{\cal L}_{0L}\nonumber\\
&&-P{\cal L}_{1c}{\cal L}_{0}^{\circ -1}
{\cal L}_{0L}{\cal L}_{0}^{\circ -1}{\cal L}_{1c}
{\cal L}_{0}^{\circ -1}{\cal L}_{0L}   \\
{\cal L}_{cL}&=&P{\cal L}_{1L}{\cal L}_{0}^{\circ -1}{\cal L}_{1c}{\cal L}_{0}^{\circ -1}{\cal L}_{0L}\nonumber\\
&&+P{\cal L}_{1c}{\cal L}_{0}^{\circ -1}{\cal L}_{0L}{\cal L}_{0}^{\circ -1}{\cal L}_{1L}\nonumber\\
&&+P{\cal L}_{1c}{\cal L}_{0}^{\circ -1}{\cal L}_{1L}{\cal L}_{0}^{\circ -1}{\cal L}_{0L}
\end{eqnarray}
where the terms which trivially vanish have been omitted.
Here,
\begin{eqnarray}
{\cal L}_{0L}\rho=-\frac{i}{\hbar}[H_{0L},\rho]
\end{eqnarray}
describe the laser-atom interaction at zero order in $\eta$ where $H_{0L}$ is defined in Eq.~(\ref{H0L}),
\begin{eqnarray}
{\cal L}_{1L}\rho&=&-\eta\frac{i}{\hbar}[(b\da+b)V_L,\rho]\\
{\cal L}_{1c}\rho&=&-\eta\frac{i}{\hbar}[(b\da+b)V_c,\rho]
\end{eqnarray}
describes respectively the laser-atom and cavity-atom interaction at first order in $\eta$ where $V_L$ and $V_c$ are defined in
Eqs.~(\ref{1stlaser})-(\ref{1stcavity}). The diffusion due to carrier excitation is given by
\begin{eqnarray}
{\cal L}_{\rm diff}&=&P{\cal L}_{2s}{\cal L}_{0}^{\circ -1}{\cal L}_{0L}{\cal L}_{0}^{\circ -1}{\cal L}_{0L}\\
\end{eqnarray}
where
\begin{eqnarray}
{\cal L}_{2s}\rho&=&\frac{\gamma\alpha}{2}\sigma\left[2(b+b\da)\rho(b+b\da)\right.\nonumber\\
&&\left.-(b+b\da)^2\rho-\rho(b+b\da)^2\right]\sigma\da
\end{eqnarray}
is the Liouvillian at second order in $\eta$ for the atomic spontaneous emission.

Tracing over the internal degree of freedom we obtain an equation
for the density matrix $\mu={\rm Tr}_I\pg{P\rho}$ for the center-of-mass
variables,
\begin{eqnarray}
Tr_I\pg{{\cal L}_{\rm diff}P\rho}&=&
\eta^2[ D(b\mu b\da-b\da b\mu+b\da\mu b-bb\da\mu)+{\rm H.c.}]\nonumber\\
Tr_I\pg{{\cal L}_{j}P\rho}&=&\eta^2[S_j(\nu)(b\mu b\da-b\da b\mu)\nonumber\\
&&+S_j(-\nu)(b\da\mu b-bb\da\mu)+{\rm H.c.}]
\end{eqnarray}
with $j=\{L,c,cL\}$. The coefficients $D$ and $S_j$ are defined as
\begin{eqnarray}
\label{weakdriveDiff}
D=\frac{\gamma\alpha}{2}\mbox{Tr}_I\{\sigma\da \sigma \mathcal{L}_{0I}^{\circ-1}{\cal
 L}_{0\Omega}\mathcal{L}_{0I}^{\circ-1}{\cal L}_{0\Omega}  \rho_{SS}^\circ\}
\end{eqnarray}
and
\begin{eqnarray}
S_L(\nu)&=&-\mbox{Tr}_I\{V_L\pt{\mathcal{L}_{0I}^\circ+i\nu}^{-1} V_L \rho_{SS}^\circ
\}\nonumber\\
S_c(\nu)&=&-\mbox{Tr}_I\{V_{c} \pt{\mathcal{L}_{0I}^\circ+i\nu}^{-1}
\left[V_{c}\mathcal{L}_{0I}^{\circ-1}\mathcal{L}_{0L}\right.\nonumber\\
&&\left.+ \mathcal{L}_{0L}\pt{\mathcal{L}_{0I}^\circ+i\nu}^{-1}V_{c}  \right]
\mathcal{L}_{0I}^{\circ-1}   \mathcal{L}_{0L}\rho_{st}^\circ   \}\nonumber\\
S_{cL}(\nu)&=&\mbox{Tr}_I\left\{V_{L}\pt{\mathcal{L}_{0I}^\circ+i\nu}^{-1} V_{c}
\mathcal{L}_{0I}^{\circ-1}\mathcal{L}_{0L}\rho_{st}^\circ  \right.\nonumber\\
&&+V_{c} \pt{\mathcal{L}_{0I}^\circ+i\nu}^{-1}\mathcal{L}_{0L}
\pt{\mathcal{L}_{0I}^\circ+i\nu}^{-1}   V_{L} \rho_{st}^\circ \nonumber\\
&&\left.+V_{c} \pt{\mathcal{L}_{0I}^\circ+i\nu}^{-1}
V_{L}\mathcal{L}_{0I}^{\circ-1}\mathcal{L}_{0L}\rho_{st}^\circ
\right\} \label{weakdriveSnu}
\end{eqnarray}
Setting $S(\nu)= \varphi_L^2 S_L(\nu)+\varphi_c^2 S_c(\nu)+
\varphi_c\varphi_L S_{cL}(\nu)$ we find an equation of the same
form as Eq.~(\ref{externaleq}). The real part of these terms are
\begin{eqnarray}
\mbox{Re}\pg{D}&=&\gamma\alpha|{\cal T}_S|^2/2\\
\mbox{Re}\pg{S_L(\mp\nu)}&=&\left(\gamma|{\cal T}_L^{\gamma,\pm}|^2+\kappa|{\cal
 T}_L^{\kappa,\pm }|^2 \right)/2\\
\mbox{Re}\pg{S_c(\mp\nu)}&=&\left(\gamma |{\cal T}_c^{\gamma,\pm }|^2+\kappa|{\cal T}_c^{\kappa,\pm }|^2\right)/2\\
\mbox{Re}\pg{S_{cL}(\mp\nu)}&=&\left(\gamma {\cal T}_L^{\gamma,\pm
}{{\cal T}_c^{\gamma,\pm}}^*+\kappa{\cal T}_L^{\kappa,\pm} {{\cal
T}_c^{\kappa,\pm}}^* \right)/2\nonumber\\
& &+{\rm c.c.}
\end{eqnarray}
where the coefficients ${\cal T}_j$ are given explicitly in
Eqs.~(\ref{transitions})-(\ref{transitions:F}). Finally, heating
and cooling rates are given by $A_{\pm}=2\mbox{Re}\{S(\mp\nu)+D\}$
and their explicit dependence on the physical parameters are
reported in Eq.~(\ref{app:rateweak}).

\section{Limit of small cavity loss rate}\label{app:smallk}

In this appendix we discuss the derivation of the rate equation,
in the limit of small cavity loss and $\delta_c=0$, by making no assumption over the laser Rabi frequency $\Omega$, which may saturate the atomic transition. The results we obtain in this appendix are valid provided that $\tilde g\neq 0$.

In order to derive the rate equation for the atomic motion
we closely follow the general approach
described in section \ref{rateeq}, where here we expand at second
order in $\eta$ and at first order in $\kappa$. The limit of
applications of the perturbative expansion are found after
identifying the smallest rate determining the internal dynamics
for $\kappa=\delta_c=0$. This is the width of the narrow resonance
$\gamma_-$, which for sufficiently large values of $|\Delta|$ takes the form
\begin{eqnarray}
\gamma_-\sim\frac{\gamma}{4}\pt{1-\frac{|\Delta|}{\sqrt{\Delta^2+4\tilde
g^2}}}
\end{eqnarray}
Therefore, an expansion in $\kappa$ and $\eta$ is possible
provided that $\kappa\ll\gamma_-$ and $\eta\varphi_c \tilde{g},~\eta\varphi_L\Omega
\ll\gamma_-$. In this regime, the
internal dynamics at zero order are described by the Liouvillian
\begin{eqnarray}
\mathcal{L}_{00I}
&=&\mathcal{L}_{0I}\Bigl|_{\kappa=0,\delta_c=0}\\
&=&\frac{1}{{\rm i}\hbar}[H_{\rm at}+H_{0{\rm at-cav}}+H_{0L},\rho]+{\cal L}_{0s}\rho\nonumber
\end{eqnarray}
and the steady state, which is
solution of $\mathcal{L}_{00I}\rho_{\rm 0St}=0$, is given by
Eq.~(\ref{Steady:smallk}). The superoperator at zero order in $\eta$ and $\kappa$ is given by
\begin{eqnarray}
{\cal L}_{00}=\mathcal{L}_{00I}+\mathcal{L}_{0E}
\end{eqnarray}

and the corresponding projector over the eigenspace at eigenvalue zero is
$P\rho=\kb{g,\beta_c}{g,\beta_c}\otimes\sum_n|n\rangle\langle
n|\mbox{Tr}_I\{\langle n|\rho|n\rangle\}$. The dynamics of this subspace at the
lowest relevant order in $\eta$ and $\kappa/\gamma_-$ are
$P\dot\rho={\cal L}_PP\rho$ where
\begin{eqnarray}
{\cal L}_P
=\sum_{n=0,m=0}^\infty\eta^n\frac{\kappa^m}{\gamma_-^m}{\cal L}_{n,m}
\end{eqnarray}
At lowest order it has the form
\begin{eqnarray}
{\cal L}_P
\simeq\eta^2{\cal L}_{2,0}+\eta^2\frac{\kappa}{\gamma_-}{\cal L}_{2,1}
\end{eqnarray}
where
\begin{eqnarray}
\eta^2{\cal L}_{2,0}&=&-P{\cal L}_{1}{\cal L}_{00}^{-1}{\cal L}_{1}\\
\eta^2\frac{\kappa}{\gamma_-}{\cal L}_{2,1}&=&
P{\cal L}_{1}{\cal L}_{00}^{-1}{\cal L}_{1}{\cal L}_{00}^{-1}{\cal K}\nonumber\\
&&+P{\cal L}_{1}{\cal L}_{00}^{-1}{\cal K}{\cal L}_{00}^{-1}{\cal L}_{1}
\end{eqnarray}
with
\begin{eqnarray}
{\cal L}_{1}=-\eta\frac{i}{\hbar}[(b\da+b)V_1,\rho]
\end{eqnarray}
Liouvillian at first order in $\eta$ for the coupling between atom and electromagnetic field, and $V_1$ given by Eqs.~(\ref{V1}). Note that we have omitted to write the terms which trivially vanish.
Tracing over the internal degree of freedom we obtain the equation
$$\dot\mu={\rm Tr}_I\pg{{\cal
L}_P\rho_{0SS}\otimes\mu}$$ for the center-of-mass variables
density matrix $\mu=Tr_I\pg{P\rho}$, whereby
\begin{eqnarray}
{\rm Tr}_I\pg{{\cal L}_{2,\ell}P\rho}&=&\eta^2[S_{\ell}(\nu)(b\mu b\da-b\da b\mu)\nonumber\\
&&+S_{\ell}(-\nu)(b\da\mu b-bb\da\mu)+{\rm H.c.}]
\end{eqnarray}
Here the index $\ell=\{0,1\}$ indicate the order of the expansion
in $\kappa$. From this equation  we can identify the coefficients
of Eq.~(\ref{externaleq}), and thus
$$D=0$$ and
$$S(\nu)=S_0(\nu)+S_1(\nu)$$ where
\begin{eqnarray}\label{Sk}
S_{0}(\nu)&=&-\mbox{Tr}_I\{V_{1} \pt{\mathcal{L}_{00I}+{\rm i}\nu}^{-1}  V_{1}
\rho_{0SS} \} \nonumber\\
S_{1}(\nu) &=&\mbox{Tr}_I\{V_{1}\pt{\mathcal{L}_{00I}+{\rm i}\nu}^{-1} \nonumber\\
& &\left[ V_{1} \mathcal{L}_{00I}^{-1}\mathcal{K}
+\mathcal{K}\pt{\mathcal{L}_{00I}+{\rm i}\nu}^{-1}   V_{1}\right]\rho_{0SS
}\}\nonumber\\
\end{eqnarray}
Note that $D={\rm o}(\kappa^2)$, as in this system the population
of the atomic excited state grows quadratically with
$\kappa$~\cite{Zippilli04a,Zippilli04b}.

Heating and cooling rates are found from the relation
$A_\pm=2\mbox{Re}\{S_0(\mp\nu)+S_1(\mp\nu)\}$, and take the form
\begin{eqnarray}\label{app:ratesmallk}
A_{\pm}=\Omega^2 {\cal A}_\pm\left(\varphi_L^2 +\varphi_c^2+
\xi_{\kappa}^\pm\right)
\end{eqnarray}
whereby $|\xi_{\kappa}^\pm|\ll 1$. In particular,
\begin{eqnarray}
{\cal A}_\pm =\frac{ \nu^2\gamma}{[\nu(\nu\mp\Delta)-\tilde
g^2]^2+\nu^2\gamma^2/4}
\end{eqnarray}
with
\begin{eqnarray}
\label{a:pm:1:k}
\xi_{\kappa}^\pm&=&
  \frac{\kappa ^2}{\nu ^2}C_1 \left(1-\frac{ \gamma }{2}{\cal A}_\pm\right)
  (\varphi_L^2 +\varphi_c^2)\nonumber\\
&& -\frac{\varphi_c^2 }{2 C_1}+\frac{\kappa}{\nu }
\left(\frac{\Delta  \nu }{\tilde g^2}\mp1\right)
  \varphi_L \varphi_c
\end{eqnarray}

Result~(\ref{app:ratesmallk}) coincides with the one obtained from
expanding Eq. (\ref{app:rateweak}) to the first order in $\kappa$
and with $\delta_c=0$, from which the results in Eqs.~(\ref{W:CIT}) and~(\ref{n:K}) have been obtained. Nevertheless, in
deriving rates~(\ref{app:ratesmallk}) we have made no assumption on the
strength of the laser intensity.
\end{appendix}


\begin{thebibliography}{99}
\bibitem{ReviewCooling}
C.\ Cohen-Tannoudij, "Atomic motion in laser light" in {\it Fundamental Systems
in Quantum Optics}, Les Houches Summer School Proceedings, Vol.\ 53, p.\ 1-164,
J.\ Dalibard, J.-M.\ Raymond and J.\ Zinn-Justin, eds.\ (North Holland, Amsterdam, 1992).
\bibitem{Vuletic00}
V.\ Vuletic and S.\ Chu, Phys.\ Rev.\ Lett.\ {\bf 84}, 3787
(2000).
\bibitem{Domokos03}
P.\ Domokos and H.\ Ritsch, J.\ Opt.\ Soc.\ Am.\ B {\bf 20}, 1098
(2003).
\bibitem{Pinkse00}
P.W.H.\ Pinkse, T.\ Fisher, P.\ Maunz, and G.\ Rempe, Nature (London) {\bf
  404}, 365 (2000).
\bibitem{Hood00}
C.J.\ Hood, T.W.\ Lynn, A.C.\ Doherty, and H.J.\ Kimble, Science {\bf
  287}, 1447 (2000).
\bibitem{Rempe04}
P.\ Maunz, T.\ Puppe, I.\ Schuster, N.\ Syassen, P.W.H.\ Pinkse,
G.\ Rempe, Nature {\bf 428}, 50 (2004).
\bibitem{KimbleFORT03}
J.R.\ Buck, A.D.\ Boozer, A.\ Kuzmich, H.C.\ N\"agerl, D.M.\
Stamper-Kurn, H.J.\ Kimble, Phys.\ Rev.\ Lett.\ {\bf 90}, 133602
(2003).
\bibitem{Vuletic03}
H.W.\ Chan, A.T.\ Black, V.\ Vuletic, Phys.\ Rev.\ Lett.\ {\bf
90}, 063003 (2003); A.T.\ Black, H.W.\ Chan, V.\ Vuletic A.T.\
Black, Phys.\ Rev.\ Lett.\ {\bf 91}, 203001 (2003).
\bibitem{Zimmermann03}
D.\ Kruse, C.\ von Cube, C.\ Zimmermann, Ph.W.\ Courteille, Phys.\
Rev.\ Lett.\ {\bf 91}, 183601 (2003);
C.\ von Cube, S.\ Slama, D.\ Kruse, C.\ Zimmermann, Ph.W.\ Courteille, G.R.M.\ Robb, N.\ Piovella, and R.\ Bonifacio, Phys.\ Rev.\ Lett.\ {\bf 93}, 083601 (2004);
S.\ Slama, C.\ von Cube, B.\ Deh, A.\ Ludewig, C.\ Zimmermann, and Ph.W.\ Courteille,
Phys.\ Rev.\ Lett.\ {\bf 94}, 193901 (2005).
\bibitem{Hemmerich03}
B.\ Nagorny, Th.\ Els\"asser, A.\ Hemmerich, Phys.\ Rev.\ Lett.\
{\bf 91}, 153003 (2003).
\bibitem{Buschev04}
P.\ Bushev, A.\ Wilson, J.\ Eschner, C.\ Raab, F.\ Schmidt-Kaler, C.\ Becher, and R.\ Blatt,
Phys.\ Rev.\ Lett.\ {\bf 92}, 223602 (2004).
\bibitem{Kuhn05}
S.\ Nussmann, K.\ Murr, M.\ Hijlkema, B.\ Weber, A.\ Kuhn, and G.\ Rempe, 
Nat.\ Phys.\ {\bf 1}, 122 (2005).
\bibitem{Cirac95}
J.I.\ Cirac, M.\ Lewenstein, P.\ Zoller, Phys.\ Rev.\ A {\bf 51}, 1650 (1995).
\bibitem{Horak}
P.\ Horak, G.\ Hechenblaikner, K.M.\ Gheri, H. Stecher, H.\ Ritsch, Phys.\ Rev.\
 Lett.\ {\bf 79}, 4974 (1997);
G.\ Hechenblaikner, M.\ Gangl, P.\ Horak, H.\ Ritsch, Phys.\ Rev.\ A {\bf 58},
3030 (1998).
\bibitem{Vuletic01}
V.~Vuletic, H.W.~Chan, A.T.~Black, Phys.\ Rev.\ A {\bf 64}, 033405
(2001).
\bibitem{vanEnk}
S.J.\ van Enk, J.\ McKeever, H.J.\ Kimble, and J.\ Ye,
Phys.\ Rev.\ A {\bf 64}, 013407 (2001)
\bibitem{Domokos02}
P.\ Domokos and H.\ Ritsch, Phys.\ Rev.\ Lett.\ {\bf 89}, 253003
(2002).
\bibitem{Domokos02b}
P.\ Domokos, Th.\ Salzburger, and H.\ Ritsch,
Phys.\ Rev.\ A {\bf 66}, 043406 (2002)
\bibitem{Domokos04}
P.~Domokos, A.~Vukics, H.~Ritsch, Phys.\ Rev.\ Lett.\ {\bf 92}, 103601 (2004).
\bibitem{Beige04}
A.\ Beige, P.L.\ Knight, and G.\ Vitiello,
New J. Phys. {\bf 7}, 96 (2005).
\bibitem{Helmut04}
Th.\ Salzburger, and H.\ Ritsch, Phys.\ Rev.\ Lett.\ {\bf 93}, 063002 (2004).
\bibitem{Karim}
K.\ Murr, J.\ Phys.\ B:\ At.\ Mol.\ Opt.\ Phys.\ {\bf 36}, 2515 (2003).
\bibitem{Sauer03}
J.A.\ Sauer, K.M.\ Fortier, M.S.\ Chang, C.D.\ Hamley, M.S.\
Chapman, Phys.\ Rev.\ A {\bf 69}, 051804 (2004).
\bibitem{Guthorlein01}
G.R.\ Guth\"ohrlein, M.\ Keller, K.\ Hayasaka, W.\ Lange, and H.\ Walther,
Nature {\bf 414}, 49 (2001).
\bibitem{Keller04}
M.\ Keller, B.\ Lange, K.\ Hayasaka, W.\ Lange,  and H.\
Walther, Nature {\bf 431}, 1075 (2004).
\bibitem{Mundt02}
A.B.\ Mundt, A.\ Kreuter, C.\ Becher, D.\ Leibfried, J.\ Eschner,
F.\ Schmidt-Kaler, and R.\ Blatt, Phys.\ Rev.\ Lett.\ {\bf 89},
103001 (2002).
\bibitem{Zippilli05}
S.\ Zippilli and G.\ Morigi, Phys.\ Rev.\ Lett.\ {\bf 95},
143001 (2005).
\bibitem{Eschner03}
J.\ Eschner, G.\ Morigi, F.\ Schmidt-Kaler, R.\ Blatt, J.\ Opt.\
Soc.\ Am.\ B {\bf 20}, 1003 (2003).
\bibitem{Dalibard85}
J.\ Dalibard and C.\ Cohen-Tannoudji, J.\ Opt.\
Soc.\ Am.\ B {\bf 11}, 1707 (1985).
\bibitem{Morigi00}
G.\ Morigi, J.\ Eschner, C.H.\ Keitel, Phys.\ Rev.\ Lett. {\bf
85}, 4458 (2000); C.F.~Roos, D.~Leibfried, A.~Mundt,
F.~Schmidt-Kaler, J.~Eschner, and R.~Blatt, Phys.\ Rev.\ Lett.\
{\bf 85}, 5547 (2000); F.~Schmidt-Kaler, J.~Eschner, G.~Morigi,
C.~Roos, D.~Leibfried, A.~Mundt, and R.~Blatt, Appl.\ Phys.\ B
{\bf 73}, 807 (2001).
\bibitem{Alsing92}
P.M.\ Alsing, D.A.\ Cardimona, H.J.\ Carmichael, Phys.\ Rev.\ A {\bf 45}, 1793
(1992).
\bibitem{Zippilli04a}
S.\ Zippilli, G.\ Morigi, H.~Ritsch, Phys.\ Rev.\ Lett.\ {\bf 93} 123002 (2004).
\bibitem{Zippilli04b}
S.\ Zippilli, G.\ Morigi, H.~Ritsch, Eur.\ Phys.\ J.\ D  {\bf 31},
507 (2004).
\bibitem{Stenholm86}
S.\ Stenholm, Rev.\ Mod.\ Phys.\ {\bf 58}, 699 (1986).
\bibitem{Javanainen84}
J.~Javanainen, M.~Lindberg, S.~Stenholm,
J. Opt. Soc. Am. {\bf B1}, 111 (1984).
\bibitem{Cirac92}
J.I.\ Cirac, R.\ Blatt, P.\ Zoller, W.D.\ Phillips, Phys.\ Rev.\ A {\bf 46}, 266
8 (1992).
\bibitem{Morigi03}
G.\ Morigi, Phys.\ Rev.\ A {\bf 67}, 033402 (2003).
\bibitem{Bienert04}
M.~Bienert, W.~Merkel, G.~Morigi, Phys.\ Rev.\ A {\bf 69}, 013405
(2004).
\bibitem{Englert}
H.J.\ Briegel and B.-G.\ Englert, Phys.\ Rev. A {\bf 47}, 3311
(1993); For a review, see B.-G.\ Englert and G.\ Morigi, in {\it
Coherent Evolution in Noisy Environments}, Lecture Notes in
Physics {\bf 611}, p.\ 55, ed.\ by A. Buchleitner, K. Hornberger
(Springer Verlag, Berlin-Heidelberg-New York 2002), and references
therein.
\bibitem{Nienhuis91}
G.\ Nienhuis, P.\ van der Straten, and S–Q.\ Shang,
Phys.\ Rev.\ A {\bf 44}, 462 (1991).
\bibitem{Kimble94}
H.J.\ Kimble, in {\it Cavity Quantum Electrodynamics}, p. 203,
ed.\ by P.R.\ Berman, Academic Press (New York, 1994).
\bibitem{Evers}
J.\ Evers, C.H.\ Keitel, Europhys.\ Lett.\ {\bf
68}, 370 (2004).
\bibitem{Footnote}
{Ideally, for $\delta_c=-\nu$ we should
take $\Delta\to\infty$. Here, we have taken a large but finite
value, consistent with the expansion in $1/\Delta$ presented in
Sec.~\ref{Sec:4:Bad}.}
\bibitem{Marzoli94}
I.\ Marzoli, J.I.\ Cirac, R.\ Blatt, P.\ Zoller, Phys.\ Rev.\ A
{\bf 49}, 2771 (1994).
\bibitem{Rice04}
J.\ Leach and P.R.\ Rice, Phys.\ Rev.\ Lett.\ {\bf 93}, 103601 (2004).
\bibitem{Zippilli05b}
S.\ Zippilli, G.\ Morigi, W.P.\ Schleich, (unpublished).
\end{thebibliography}
\end{document}